\documentclass[twocolumn,3p,times,procedia]{elsarticle}

\usepackage{ecrc}
\usepackage{graphicx}
\usepackage{balance}
\usepackage{array,threeparttable}
\usepackage{url}
\volume{00}

\firstpage{1}
\runauth{}
\jid{procs}
\CopyrightLine{2026}{Published by Elsevier Ltd.}
\usepackage{amssymb}
\usepackage{amsmath}
\usepackage{multicol}
\usepackage{algorithm}
\usepackage{algpseudocode}
\usepackage[figuresright]{rotating}
\usepackage{bm}
\usepackage{caption}
\usepackage{subfigure} 
\usepackage{titlesec}
\titleformat*{\section}{\small \bf}
\titleformat*{\subsection}{\small \em}
\titleformat*{\subsubsection}{\small \em}

\usepackage{fancyhdr}
\fancyheadoffset[RO,EL]{0pt}
\usepackage{geometry}
\begin{document}\small
\begin{frontmatter}




\dochead{}
\title{
\begin{flushleft}
{\LARGE Reinforcement Learning-Based Secure Near-field Directional Modulation Enhanced by Rotatable RIS}
\end{flushleft}
}
 %

\author[]{ \leftline {Yongqiang Li$^a$, Feng Shu$^*$$^{a,b}$, Shaofan Chen$^c$, Yuanyuan Wu$^a$, Maolin Li$^a$,  Zhen Chen$^d$, } \leftline{Hao Jiang$^{e,f}$,Jiangzhou Wang$^e$}}

\address{ \leftline {$^a$School of Information and Communication Engineering, Hainan University, Haikou 570228, China}  \leftline{$^b$School of Electronic and Optical Engineering, Nanjing University of Science and Technology, Nanjing 210094, China.}
	\leftline{$^c$School of Computer Science and Technology, Hainan University, Haikou 570228, China}
	\leftline{College of Cyber Security, Jinan University, Guangzhou, 510632, China}
   \leftline{$^e$National Mobile Communications Research
  Laboratory, Southeast University, Nanjing 210096, China.}
   \leftline{$^f$School of Artificial Intelligence, Nanjing University of Information Science and Technology, Nanjing 210044, China.}
}

\cortext[]{Feng Shu (Corresponding author) Feng Shu is with the School of Information and Communication Engineering, Hainan University, Haikou 570228, China, and also with the School of Electronic and Optical Engineering, Nanjing University of Science and Technology, Nanjing 210094, China. (e-mail: shufeng0101@163.com).}

\fntext[]{Shaofan Chen is with the School of Computer Science and Technology, Hainan University, Haikou, 570228, China. (e-mail:990588 @hainanu.edu.cn).}

\fntext[]{Yuanyuan Wu and Maolin Li are with the School of Information and Communication Engineering, Hainan University, Haikou, 570228, China. (e-mail:995042@hainanu.edu.cn; limaolin0302@163.com).}

\fntext[]{Zhen Chen is with the College of Cyber Security, Jinan University, Guangzhou, 510632, China (e-mail: chenz.scut@gmail.com).}

\fntext[]{Hao Jiang is with the National Mobile Communications Research Lab
oratory, Southeast University, Nanjing 210096, China, and also with the
School of Artificial Intelligence, Nanjing University of Information Science
and Technology, Nanjing 210044, China (e-mail: jianghao@nuist.edu.cn).}

\fntext[]{Jiangzhou Wang is with the National Mobile Communications Research
Laboratory, Southeast University, Nanjing 210096, China (e-mail:
j.z.wang@kent.ac.uk).}

\begin{abstract}
This paper investigates secure Directional Modulation (DM) design enhanced by a rotatable active Reconfigurable Intelligent Surface (RIS). In conventional RIS-assisted DM networks, the security performance gain is limited due to the multiplicative path loss introduced by the RIS reflection path. To address this challenge, a Secrecy Rate (SR) maximization problem is formulated, subject to constraints including the eavesdropper's Direction Of Arrival (DOA) estimation performance, transmit power, rotatable range, and maximum reflection amplitude of the RIS elements. To solve this non-convex optimization problem, three algorithms are proposed: a multi-stream null-space projection and leakage-based method, an enhanced leakage-based method, and an optimization scheme based on the Distributed Soft Actor-Critic with Three refinements (DSAC-T). Simulation results validate the effectiveness of the proposed algorithms. A performance trade-off is observed between eavesdropper's DOA estimation accuracy and the achievable SR. The security enhancement provided by the RIS is more significant in systems equipped with a small number of antennas. By optimizing the orientation of the RIS, a 52.6\% improvement in SR performance can be achieved.

\end{abstract}

\begin{keyword}

Directional modulation \sep Secure communication \sep  Rotatable RIS\sep DSAC-T\sep CRLB


\end{keyword}

\end{frontmatter}


\section{Introduction}
	With the advancement of 6G technology and standardization, a number of promising solutions have been explored to support future wireless communication networks. Various types of Reconfigurable Intelligent Surfaces (RIS) or Intelligent Reflecting Surfaces (IRS) have been studied~\cite{Xie2025}, which introduce new Degrees Of Freedom (DoF) to the system by enabling dynamic reconfiguration of the wireless propagation environment. Additionally, the intelligent tunability of antennas through mechanical and electronic means has been investigated in communication systems, including Fluid Antennas (FAs)~\cite{Wong2022,Chen2025b,Wong2023}, Movable Antennas (MAs)~\cite{Bian2026,Tang2025,Li2025c}, Pinching Antennas (PAs)~\cite{Ding2025a,Wang2025a,Xu2025}, and Rotatable Antennas (RAs)~\cite{Qu2025,Zeng2020,Zheng2025}. The intelligent control of both antennas and the wireless environment is expected to realize the vision of high-capacity and secure future wireless communications.

\subsection{Background}

Security is a critical issue in wireless communication systems. Current communication systems primarily rely on encryption techniques at the upper-layer protocols to protect data transmission security. Physical Layer Security (PLS), which leverages the inherent properties of wireless channels to enhance security, serves as a powerful complement to cryptography. According to information theory, Secrecy Rate (SR) has been investigated as a key performance metric. Leveraging the multidimensional propagation characteristics of signals in space, time, and frequency domains, a variety of physical layer security techniques have been proposed. In~\cite{Shu2024}, IRS-assisted spatial modulation was investigated, and three effective low-complexity methods were proposed to enhance SR performance. In~\cite{Man2025}, based on 1-bit phase shifters, a genetic algorithm was utilized for time-modulation design, achieving low hardware complexity and cost. In \cite{Shu2018}, a secure precise transmission technique based on random subcarrier selection was proposed, which effectively reduces circuit complexity at the receiver while achieving dual security in both angle and distance domains. Given the limited capability of a single technology to address both cost and performance challenges, signals are now being designed and integrated across multiple dimensions to enhance overall system security. By jointly designing across frequency and time domains, an anti-jamming modulation scheme was proposed in \cite{Fang2024}. By partitioning the array into multiple sub-arrays, a pseudo-random modulation technique was employed in \cite{Li2025} to achieve high-precision and independent control of multiple beams. Effective combinations of technologies are being explored and investigated to achieve higher security performance at lower cost.

Unlike traditional Fixed-position Antenna systems (FPAs), antenna elements can be electronically or mechanically controlled to move, leading to extensive research on MA technology. By endowing antennas with mobility, more flexible spatial configurations can be achieved, enabling easy switching to special array structures such as coprime arrays~\cite{Chen2025} and sparse arrays~\cite{Zhuang2024}. This enhances system performance in complex wireless communication environments.
In comparison with RIS technology, MAs design transmission waveforms more favorable to legitimate users by jointly optimizing antenna positions and precoding matrices. Active RIS~\cite{Shi2024}, on the other hand, adjusts both amplitude and phase of incident signals, providing additional distinguishable and controllable paths for signal enhancement, interference suppression, and multiplexing gain.
In essence, MAs can be viewed as holistically improving channel quality, whereas RIS enhances system performance by locally refining channel conditions. 
Distinct from their physical implementation, FAs are considered a technology analogous to MAs. In~\cite{Chen2025a}, covert transmission in FA systems was studied under two scenarios: one where the Eavesdropper's (Eve's) location is known, and another where it is unknown~\cite{Zhang2025a}. In both cases, the probability of being eavesdropped was significantly reduced. For both scenarios, a low-complexity beamforming algorithm design based on MA arrays was proposed in \cite{Li2025a}. It was verified that, given a sufficient range of movement, MAs can achieve higher security performance compared to traditional FPA systems. With the increase in array aperture, some users may fall within the near-field region~\cite{Li2026}. MAs can achieve superior beam focusing performance. Compared to FPA systems, secure transmission can still be realized even when an Eve located in the same direction is in closer proximity to the legitimate user (Bob)~\cite{Ma2025}. From a physical layer security perspective, reference \cite{Li2025b} discussed the performance of both macro- and micro-scale MAs, revealing the potential of hybrid micro-macro mobility in 6G networks. A secure transmission scheme based on MAs was proposed in~\cite{Hu2024}. This scheme does not require estimating the Eve’s instantaneous channel. Compared to FPA systems, it can achieve improved secrecy outage probability. MA-enhanced secure beamforming for full-duplex multi-user communication was investigated in~\cite{Ding2025}. The MA technique can also achieve superior security performance in integrated sensing and communication systems~\cite{Ma2025a}. A deep reinforcement learning approach was utilized in \cite{LeHung2025} to jointly optimize the transmit beamforming and MA movement, achieving a SR gain of $40\%$.

However, the performance gain achieved by the MA is limited due to the restricted movable range~\cite{Li2025a,Li2026}. It is noteworthy that while both MAs and RISs can reconfigure the wireless propagation environment, optimized passive beamforming of RIS may diminish the performance gains achieved by MAs compared to FPA systems~\cite{Wei2025}. To further enhance the system DoF, RAs have emerged as a research hotspot~\cite{Zhang2025,Xiong2025,Xiong2025a}.
Each antenna element, characterized by a specific main lobe width, can be independently rotated. Mathematically, the main lobe width is flexibly modeled by adjusting the directivity factor, allowing the theoretical framework to span from isotropic to highly directional radiation patterns. In \cite{Zheng2025a}, two low-complexity algorithms were employed to enhance the array gain, demonstrating performance superior to that of FPA systems. Furthermore, by steering the orientation of the antennas, interference can be significantly mitigated, thereby validating the interference suppression capability of RAs~\cite{Zheng2025b}.
Secure wireless communication enhanced by RAs was investigated in~\cite{Dai2025}. By leveraging the generalized Rayleigh quotient and the successive convex approximation algorithm for joint beamforming design and deflection angle optimization, the proposed scheme achieves superior performance compared to FPA systems, isotropic antenna systems, and random deflection angle schemes.
It is noteworthy that a Six-dimensional Movable Antenna (6DMA) architecture has been proposed, enabling multiple sub-arrays to move and rotate in three-dimensional space, thereby demonstrating coverage enhancement and performance improvement capabilities. Continuous and discrete 6DMA movements were studied in \cite{Shao2025a}
and \cite{Shao2025}, respectively, both showing significant gains in transmission rate.
6DMA was further investigated in \cite{Shao2025b}, which presented prototype experiments for both a localizable 6DMA with fixed antenna rotation and a rotatable 6DMA with fixed antenna positions. The study also outlined key applications and challenges related to its architecture, antenna position and rotation optimization~\cite{Shao2025d, Wang2025,Pi2025}, channel estimation~\cite{Shao2025c}, and system design from both communication and sensing perspectives~\cite{Hua2025}.
A RIS equipped with a large number of adjustable elements can also be rotated to enhance system performance. A RIS with both mobility and rotational capability was proposed in~\cite{Lu2025} to enhance coverage. Furthermore, the PAs are also regarded as a type of flexible antenna systems. Activated by applying small dielectric particles onto a dielectric waveguide, it features low cost and flexible deployment, aiming to overcome obstacle blockage by creating effective Line-Of-Sight (LoS) paths~\cite{Wang2025c}. In \cite{Wang2025b}, a coalition game-based algorithm was proposed to optimize the combination of activated antennas. By degrading the signal quality of eavesdroppers, the SR can be significantly improved.

\subsection{Motivations and Contributions}
However, although various types of RIS have been demonstrated to enhance security, research on rotatable RIS remains in its infancy. Furthermore, most existing works assume perfect Channel State Information (CSI) or statistical CSI error models. However, in practical scenarios, the angle estimation error of the target is stochastic. In this paper, we investigate the secrecy performance of a rotatable RIS-aided network. To ensure instantaneous secrecy performance, a framework involving sequential Direction-Of-Arrival (DOA) estimation followed by secure beamforming design is proposed. The main contributions are summarized as follows:
	\begin{itemize}
	
	\item A rotatable active RIS-assisted DM network is established. SR maximization problem is formulated under constraints on the Cramér–Rao Lower Bound (CRLB), transmit power, rotatable range, and maximum reflection amplitude of the RIS elements. Distinct from conventional approaches that treat DOA estimation and DM design separately, the CRLB is incorporated into the DM design as a metric for DOA estimation performance.
	
	\item Two low-complexity algorithms and a learning-based scheme are respectively proposed to solve the SR maximization problem. Firstly, a method based on multi-stream null-space projection and leakage is introduced. Secondly, based on leakage theory, an enhanced leakage-based method is presented. Finally, an optimization scheme based on DSAC-T is proposed, aiming to optimize the RIS orientation to further improve SR performance.
	
	\item The simulation results validate the effectiveness of the proposed algorithms. Due to multiplicative path loss, the security of the data stream transmitted via the RIS path is lower than that of the LoS path. Optimizing the orientation of the RIS can enhance the SR, and the security improvement offered by the RIS reflection path is more pronounced, especially in DM systems equipped with a small number of antennas. By increasing the maximum reflection amplitude of the RIS elements, the security level of the RIS reflection path can be elevated to a level comparable to that of the LoS path. Furthermore, compared to Algorithm 2, the DSAC-T-based optimization scheme can achieve a 52.6\% improvement in SR performance.
\end{itemize}	
	\subsection{Organization and Notation}

The remaining part of the paper is as follows. Sec. \ref{sec:2} is the system model and problem formulation. In Sec. \ref{sec:3},  three algorithms are presented. In Sec. \ref{sec:4}, simulation results are provided. The conclusion is illustrated in Sec. \ref{sec:5}.

Notations: x, $\mathbf{x}$, and $\mathbf{X}$ represent a scalar, a vector, and a matrix, respectively. $\mathcal{X}$ denotes a set. The space of  $M\times N$ complex matrices is denoted by $\mathbb{C}^{M\times N}$. The modulus, Euclidean norm, and infinity norm are represented by $|\ |$, $\|\ \|$, and $\|\ \|_{\inf}$, respectively. Superscripts $[\ ]^T$, $[\ ]^*$, and $[\ ]^H$ stand for transpose, complex conjugate, and conjugate transpose (Hermitian) operators, respectively.$\mathbb{E}\{\ \}$ stands for expectation.  The operator $\text{diag}(\ )$ forms a diagonal matrix from its argument The operator $[\ ]^+$ returns the argument if it is non‑negative, and zero otherwise.
\section{System model and problem formulation}\label{sec:2}
\subsection{System model}
\begin{figure}
	\centering
	\includegraphics[width=0.45\textwidth, trim = 10 5 2 0,clip]{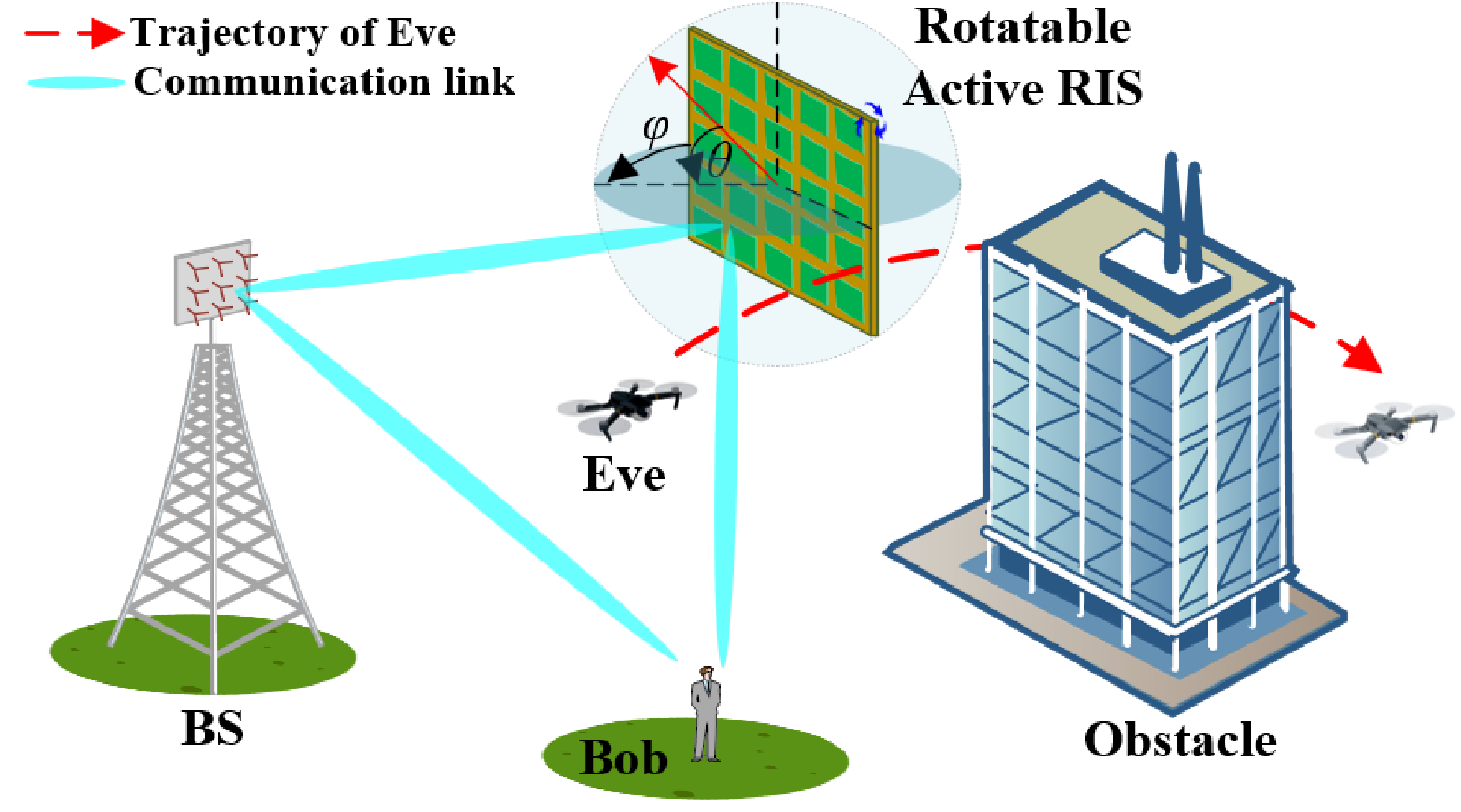}\\
	\caption{Rotatable active RIS-enhanced secure communication model.}\label{fig:1}
\end{figure}
A secure communication model is considered, as illustrated in Fig. \ref{fig:1}, consisting of a Base Station (BS), a Bob, and an Eve. The BS transmits confidential signals to Bob via both a LoS path and an active RIS-reflected path, while Eve attempts to eavesdrop on the transmitted information. Eve is modeled as a low-altitude uncrewed aerial vehicle. The objective is to leverage the rotational capability of the RIS as a potential DoF to enhance security. A spherical coordinate system is adopted with the center of the RIS serving as the origin. The RIS is a planar array composed of $M=M_hM_v$ units. The BS is equipped with a fixed uniform planar antenna array with $N=N_hN_v$ antennas. Here, $N_h (M_h)$ and $N_v (M_v)$ denote the number of BS antenna elements (RIS reflecting elements) in the horizontal and vertical directions, respectively.

The heights of the BS and the RIS relative to Bob are $H_b$ and $H_r$, respectively. The elevation and azimuth angles from the BS to Bob and Eve are $\theta_{ab}$ and $\phi_{ab}$, and $\theta_{t,ae}$ and $\phi_{t,ae}$, respectively. Relative to the coordinate origin (the center of RIS), the elevation and azimuth angles corresponding to the BS are represented by $\theta_{a}$ and $\phi_{a}$, and those corresponding to Bob can be represented by $\theta_{b}$ and $\phi_{b}$, respectively. Let $\theta_{t,e}$ and $\phi_{t,e}$ denote the elevation angle and the azimuth angle form RIS to Eve at time slot $t$, respectively. The positions of the BS reference antenna and Bob can be represented by $\mathbf{a} = [r_{a,x}, r_{a,y}, r_{a,z}]^T$ and $\mathbf{b}_t = [r_{b,x}, r_{b,y}, r_{b,z}]^T$, respectively. The continuous trajectory of Eve is sampled into a set of discrete positions partitioned over $T$ ($t \in\mathcal{T}= \{1, \ldots, T\}$) time slots. We assume that the BS estimates Eve's instantaneous position at uniform intervals, with each time slot having equal duration. The sampled trajectory of Eve can be represented by $\mathbf{u}_{t,e} = [r_{t,e,x}, r_{t,e,y}, r_{t,e,z}]^T$.
The RIS may undergo different rotations at different time slots. For the $t$-th time slot, the position of the $m$-th ($m\in\mathcal M = \{1,2,...,M\}$) unit can be represented by $\mathbf{i}_{t,m} = [x_{t,m}, y_{t,m}, z_{t,m}]^T$. 

For clarity, the position representations of RIS units and all antenna elements are provided. Specifically, the normal vector of the RIS and the two orthogonal basis vectors perpendicular to it can be represented by $\mathbf{k}=[1,0,0]^T$, $\mathbf{k}_h=[0,1,0]^T$, and $\mathbf{k}_v=[0,0,1]^T$, respectively. $\mathbf{k}$, $\mathbf{k}_h$, and $\mathbf{k}_v$ denote the unit basis vectors of the local coordinate system centered at the RIS origin, respectively. Then, the rotated unit coordinate system can be represented by
\begin{align}
	&\tilde{\mathbf{k}}(\alpha,\beta)=[\cos\alpha\cos\beta,\cos\alpha\sin\beta,\sin\alpha]^T\\
	&\tilde{\mathbf{k}}_h(\alpha,\beta)=[-\sin\beta,\cos\beta,0]^T\\
	&\tilde{\mathbf{k}}_v(\alpha,\beta)=[-\sin\alpha\cos\beta,-\sin\alpha\sin\beta,\cos\alpha]^T
\end{align}
where $\alpha\in[-\pi/2,\pi/2]$ and $\beta\in[0,2\pi]$ denote the elevation angle and azimuth angle of the rotation in the global spherical coordinate system, respectively.
The distance of the $m$-th RIS unit from the origin can be represented by
\begin{align}
	r_{m,h}&=(\left\lfloor\left(m-1\right)/M_v\right\rfloor-\frac{M_h-1}{2})d_R,m\in\mathcal{M}\label{4}\\
	r_{m,v}&=(m-\left\lfloor\left(m-1\right)/M_v\right\rfloor M_v-\frac{M_v-1}{2})d_R,m\in\mathcal{M}\label{5}
\end{align}
where $d_R=\lambda/2$ characterizes the minimum distance between RIS elements. Then, we have
\begin{align}\label{eq6}
	\mathbf{i}_{t,m}(\alpha_t,\beta_t)&=r_{m,h}\tilde{\mathbf{k}}_{t,h}(\alpha_t,\beta_t)+r_{m,v}\tilde{\mathbf{k}}_{t,v}(\alpha_t,\beta_t)
\end{align}
Note that the subscript $t$ characterizes the $t$-th time slot. For brevity, the meanings of newly introduced subscripted variables will not be reiterated. Similarly, the position of the $n$-th antenna can be represented by
\begin{align}\label{eq7}
	\mathbf{a}_{n}(\theta_{a},\phi_{a})&=r_{n,h}\tilde{\mathbf{k}}_{h}(\theta_{a},\phi_{a})+r_{n,v}\tilde{\mathbf{k}}_{v}(\theta_{a},\phi_{a})+\mathbf{a}
\end{align}
where
\begin{align}
	r_{n,h}&=(\left\lfloor\left(n-1\right)/N_v\right\rfloor-\frac{N_h-1}{2})d_b,n\in\mathcal{N}\\
	r_{n,v}&=(n-\left\lfloor\left(n-1\right)/N_v\right\rfloor N_v-\frac{N_v-1}{2})d_b,n\in\mathcal{N}
\end{align}
Here, $\mathcal{N}=\{1,2,\ldots,N-1\}$ denotes the set of antennas, and $d_b$ characterizes the minimum distance between antennas.

\subsection{Channel model}
With the coordinate information of Bob, Eve, and the RIS  available at the BS, the channels can be modeled. Considering static quasi-stationary channels, the channel from BS to Bob can be represented by
\begin{align}
	\mathbf{h}_{ab}&=[\frac{\sqrt{\varsigma}}{d_{1,b}}e^{j\frac{2\pi}{\lambda}c_{1,b}},\frac{\sqrt{\varsigma}}{d_{2,b}}e^{j\frac{2\pi}{\lambda}c_{2,b}},\ldots,\frac{\sqrt{\varsigma}}{d_{N,b}}e^{j\frac{2\pi}{\lambda}c_{{N,b}}}]^T
\end{align}
where $d_{n,b}\approx\|\mathbf{a}_{1}(\theta_{a},\phi_{a})-\mathbf{b}\|$ and $\varsigma$ symbolize the distance from BS to Bob and the channel power gain, ${c}_{n,b}=r_{n,h}\cos\theta_{ab}\cos\phi_{ab}+r_{n,v}\sin\theta_{ab}$. For the $t$-th time slot, the channel from BS to Eve can be denoted by
\begin{align}
	&\mathbf{h}_{t,ae}=[\frac{\sqrt{\varsigma}}{d_{t,1,e}}e^{j\frac{2\pi}{\lambda}c_{t,1,e}},\frac{\sqrt{\varsigma}}{d_{t,2,e}}e^{j\frac{2\pi}{\lambda}c_{t,2,e}},\!\ldots,\!\frac{\sqrt{\varsigma}}{d_{t,N,e}}e^{j\frac{2\pi}{\lambda}c_{t,N,e}}]^T
\end{align}
where $d_{t,n,e}\approx\|\mathbf{a}_{1}(\theta_{a},\phi_{a})-\mathbf{u}_{t,e}\|$ symbolizes the distance from BS to Eve, ${c}_{t,n,e}=r_{n,h}\cos\theta_{t,ae}\cos\phi_{t,ae}+r_{n,v}\sin\theta_{t,ae}$. Similarly, the channels from the RIS  to Bob and Eve can be represented by
\begin{align}
	&\mathbf{h}_{t,Rb}=[\frac{\sqrt{\varsigma}}{\hat{d}_{t,1,b}}e^{-j\frac{2\pi}{\lambda}\hat{c}_{t,1,b}},\frac{\sqrt{\varsigma}}{\hat{d}_{t,2,b}}e^{-j\frac{2\pi}{\lambda}\hat{c}_{t,2,b}},\ldots,\frac{\sqrt{\varsigma}}{\hat{d}_{t,M,b}}e^{-j\frac{2\pi}{\lambda}\hat{c}_{t,M,b}}]^T\\	&\mathbf{h}_{t,Re}=[\frac{\sqrt{\varsigma}}{\hat{d}_{t,1,e}}e^{-j\frac{2\pi}{\lambda}\hat{c}_{t,1,e}},\frac{\sqrt{\varsigma}}{\hat{d}_{t,2,e}}e^{-j\frac{2\pi}{\lambda}\hat{c}_{t,2,e}},\ldots,\frac{\sqrt{\varsigma}}{\hat{d}_{t,M,e}}e^{-j\frac{2\pi}{\lambda}\hat{c}_{t,M,e,}}]^T
\end{align}
where $\hat{c}_{t,m,b}=\mathbf{i}_{t,m}(\alpha_t,\beta_t)^T\tilde{\mathbf{k}}(\theta_b,\phi_b)$ and $\hat{c}_{t,m,e}=\mathbf{i}_{t,m}(\alpha_t,\beta_t)^T\tilde{\mathbf{k}}(\theta_{t,e},\phi_{t,e})$. $\hat{d}_{t,m,b}=\|\mathbf{i}_{t,m}(\alpha_t,\beta_t)-\mathbf{b}\|$ and $\hat{d}_{t,m,e}\approx\|\mathbf{i}_{t,1}(\alpha_t,\beta_t)-\mathbf{u}_t\|$ denote the distance from RIS to Bob and Eve, respectively. For the channel from BS to RIS, we have $[\mathbf{H}_t]_{n,m}=\frac{\sqrt{\varsigma}}{d_{t,n,m}}e^{j\frac{2\pi}{\lambda}d_{t,n,m}}$, where $d_{t,n,m}=\|\mathbf{a}_{n}(\theta_{a},\phi_{a})-\mathbf{i}_{t,m}(\alpha_t,\beta_t)\|$ is the distance from the $n$-th antenna to the $m$-th RIS units. The RIS, controlled by the BS, can optimize the cascaded link by adjusting the amplitude $\varrho_{t,m}\in[1,\varrho_{\mathrm{max}}]$ and phase shift $\psi_m\in[0,2\pi]$ of the $M$ elements and attempting to compensate for the multiplicative path loss. $\varrho_{\mathrm{max}}$ is the maximum reflection coefficient. Then, the phase shift matrix $\mathbf{\Theta}_t$ can be denoted as $\mathbf{\Theta}_t=\mathrm{diag}(\bm{\upsilon}_t)=\mathrm{diag}(\varrho_{t,1}e^{j\psi_{t,1}},\varrho_{t,2}e^{j\psi_{t,2}},\ldots,\varrho_{t,M}e^{j\psi_{t,M}})$. In the far field, where plane waves are commonly assumed, the inter-element time delays vary linearly, yielding a rank-one characteristic that is independent of the array aperture. The phase difference between different array elements and the receiving point, as well as the phase difference between different RIS units and the receiving point, are functions of their respective elevation and azimuth angles. Furthermore, the path loss can be assumed to be approximately equal across all elements of the same array.

Thus, the received signal at Bob and Eve for the $t$-th time slot can be denoted as
\begin{align}
	{y}_{t,b}&=\underbrace{(\mathbf{h}_{t,Rb}^T\mathbf{\Theta}_t\mathbf{H}_t^T+\mathbf{h}_{ab}^T)(\mathbf{w}_t{s}_t+\mathbf{v}_t{\varpi}_t)}_{\text{Desired signals}}\notag\\&+\underbrace{\sum_{l=1}^{L}\mathbf{h}_{t,b,l}^T\mathbf{w}_t{s}_t}_{\text{Multipath interference}}+\underbrace{\sum_{l=1}^{L}\mathbf{h}_{t,b,l}^T\mathbf{v}_t{\varpi}_t}_{\text{AN focusing effect}}+\underbrace{\mathbf{h}_{t,Rb}^T\mathbf{\Theta}_t\mathbf{n}_{t}+z_t}_{\text{Noise}}\label{eq14}  
\end{align}
\begin{align}	{y}_{t,e}&=(\mathbf{h}_{t,Re}^T\mathbf{\Theta}_t\mathbf{H}_t^T+\mathbf{h}_{t,ae}^T)(\mathbf{w}_t{s}_t+\mathbf{v}_t{\varpi}_t)\notag\\
	&+\sum_{l=1}^{L}\mathbf{h}_{t,e,l}^T\mathbf{w}_t{s}_t+\sum_{l=1}^{L}\mathbf{h}_{t,e,l}^T\mathbf{v}_t{\varpi}_t+\mathbf{h}_{t,Re}^T\mathbf{\Theta}_t\mathbf{n}_{t}+z_t\label{eq15}
\end{align}
where $\mathbf{w}_t\in\mathbb{C}^{N\times 1}$ and ${s}_t$ denote the precoding vector and signal for the $t$-th time slot. $\mathbf{v}_t\in\mathbb{C}^{N\times 1}$ and ${\varpi}_t$ denote the precoding vector and signal corresponding to AN. $[\mathbf{w}_{t}]_n$ and $[\mathbf{v}_{t}]_n$ stand for the weights corresponding to the $n$-th antenna. Assuming that Bob and Eve experience the same level of noise, which satisfies $\mathbf{n}_{t}\sim\mathcal{CN}(0,\sigma_{I}^{2}\mathbf{I}_M)$ and ${z}_{t}\sim\mathcal{CN}(0,\sigma_{t}^{2})$. $\mathbf{h}_{t,b,l}$ and $\mathbf{h}_{t,e,l}$ denote the $l$-th NLoS channels corresponding to Bob and Eve, respectively, $l\in\mathcal{L}=\{1,2,\ldots,L\}$. For clarity, it is assumed that the number of distinguishable interference channels corresponding to Bob and Eve is equal. Note that in low-altitude communications, the LoS path serves as the primary propagation channel. In \eqref{eq14} and \eqref{eq15}, the RIS is deployed to create enhanced NLoS paths, allowing at least two data streams to be transmitted simultaneously. 
\subsection{Problem formulation}
In this section, a two-stage optimization framework for a rotatable RIS is proposed. Specifically, in the first stage, DOA estimation is performed and the RIS is steered toward the Eve to maximize the interference caused by AN. In the second stage, beamforming design is carried out to maximize the SR.


\subsubsection{Stage 1: DOA estimation of Eve and RIS orientation initialization}
Under the assumption of a static quasi-stationary channel, both Bob and RIS can collaborate with the BS to obtain precise location information. In medium-to-high Signal-to-Noise Ratio (SNR) regimes, the Multiple Signal Classification (MUSIC) algorithm can yield accurate estimates of the Eve’s azimuth and elevation angles. By leveraging the BS and IRS as known reference points, the Eve’s three-dimensional coordinates can thereby be uniquely determined. 


The received signal consists of multiple components with different power levels. To localize targets and enhance the power of desired signals, the receiving beam can be designed to amplify the power of the path from the RIS to the BS while suppressing other paths received at the BS~\cite{Chen2024}. Then, the received signals $\mathbf{x}_{t}$ for the $t$-th time slot can be expressed as
\begin{align}\label{16}
	\mathbf{x}_{t}=&\mathbf{H}_t\mathbf{\Theta}_t\mathbf{h}_{t,Re}{s}_{t,1}+\mathbf{H}_t\mathbf{\Theta}_t\mathbf{h}_{t,Rb}{s}_{t,2}+\hat{\mathbf{n}}_{t}
\end{align}
where ${s}_{t,1}$ and ${s}_{t,2}$ denote the signals corresponding to Eve and Bob, respectively. ${s}_{t,l}$ denotes the signal corresponding to the $l$-th path. $\hat{\mathbf{n}}_{t}\sim\mathcal{CN}(0,\hat{\sigma}_{t}^{2}\mathbf{I}_N)$ denotes the noise vector. 
Note that during the $t$‑th time slot, two sequential tasks are considered: DOA estimation and communication. Accordingly, the phase shift matrices corresponding to positioning and communication are denoted as $\mathbf{\Theta}_t^p=\mathrm{diag}(\bm{\upsilon}_t^p)$ and $\mathbf{\Theta}_t^c=\mathrm{diag}(\bm{\upsilon}_t^c)$, respectively, with $\mathbf{\Theta}_t\in\{\mathbf{\Theta}_t^p,\mathbf{\Theta}_t^c\}$. By collecting the RIS phase shift matrices across $P$ $(p\in\mathcal{P}=\{1,2,\ldots,P\})$ modes, the received signal can be expressed as
\begin{align}\label{17}
	\hat{\mathbf{x}}_{t}&=[({\mathbf{x}}_{t}^1)^T,({\mathbf{x}}_{t}^2)^T,\ldots,({\mathbf{x}}_{t}^P)^T]^T
\end{align}
where
\begin{align}\label{18}
	\mathbf{x}_{t}^p&=\mathbf{H}_t\mathbf{\Theta}_t^p\mathbf{h}_{t,Re}{s}_{t,1}+\mathbf{H}_t\mathbf{\Theta}_t^p\mathbf{h}_{t,Rb}{s}_{t,2}+\hat{\mathbf{n}}_{t}^p
\end{align}
$\hat{\mathbf{n}}_{t}^p\sim\mathcal{CN}(0,\hat{\sigma}_{t}^{2}\mathbf{I}_N)$ denotes the noise vector corresponding to the $p$-th RIS phase shift matrix. Since the Eve may approach from any direction and the rotation of the RIS also affects the received SINR, the interference from Bob can be minimized to improve the SINR of the RIS-reflected path. The corresponding optimization problem can be formulated as
\begin{subequations}
	\begin{align}
		\label{19}
		\text{P1:}\quad&\underset{\bm{\upsilon}_t^p}{\min}\ \|\mathbf{H}_t\mathrm{diag}(\mathbf{h}_{t,Rb})\bm{\upsilon}_t^p\|^2
		\\ 
		\mathrm{s.t.}\ &\varrho_{t,m}\leq \varrho_{\max},\forall t,m
	\end{align}
\end{subequations}
P1 can be solved using semidefinite relaxation, thereby obtaining the optimized $\mathbf{\Theta}_t^p$. By performing eigenvalue decomposition on $\mathbf{\Theta}_t^p$, the $p$-th measurement for the RIS can be denoted as
\begin{align}
	\mathbf{\Theta}_t^p=\mathbf{U}\mathbf{\Lambda}\mathbf{U}^H
\end{align}
Then, we can obtain
\begin{align}
	\bm{\upsilon}_t^p=e^{j\angle\mathbf{U}\mathbf{\Lambda}^{\tfrac{1}{2}}\hat{\bm{\upsilon}}_t^p}
\end{align}
where $\hat{\bm{\upsilon}}_t^p$ is a random vector following a complex Gaussian distribution.
Subsequently, the precise location of Eve can be obtained using the conventional MUSIC method. 

We consider using the CRLB to evaluate the DOA estimation performance. The primary objective is to estimate the elevation ($\theta_{t,e}$) and azimuth ($\phi_{t,e}$) angles of Eve. Firstly, the unknown parameters $\theta_{\text{ER}}$ and $\phi_{\text{ER}}$ are collected and denoted as 
\begin{align}
	\bm{\kappa}=[\theta_{t,e},\phi_{t,e}]^T
\end{align}
Correspondingly, the probability density function of $\hat{\mathbf{x}}_{t}$ can be denoted as
\begin{align}
	f(\hat{\mathbf{x}}_{t};\bm{\kappa})=\frac{1}{\pi^{NP}\det(\bm{\Sigma}_t)}e^{-[\hat{\mathbf{x}}_{t}-\bm{\mu}_t]^H\bm{\Sigma_t}^{-1}[\hat{\mathbf{x}}_{t}-\bm{\mu}_t]}
\end{align}
where
\begin{align}
	\bm{\Sigma}_t&=\hat{\sigma}_t^2\mathbf{I}_{NP}\\
	\bm{\mu}_t&=[\hat{\mathbf{s}}_t^T(\hat{\mathbf{H}}_t^1)^T,\hat{\mathbf{s}}_t^T(\hat{\mathbf{H}}_t^2)^T,\ldots,\hat{\mathbf{s}}_t^T(\hat{\mathbf{H}}_t^P)^T]^T
\end{align}
Then, with the likelihood function $\ln f(\hat{\mathbf{x}}_{t};\bm{\kappa})$, the Fisher information matrix (FIM) can be expressed as
\begin{align}
	\boldsymbol{F}&=\begin{bmatrix}{\Omega}_{1,1}&{\Omega}_{1,2}\\
		{\Omega}_{2,1}&{\Omega}_{2,2}
	\end{bmatrix}\notag\\
	&=2\hat{\sigma}_t^{-2}\begin{bmatrix}{\mathbf{b}_1^H\mathbf{G}^H\mathbf{G}\mathbf{b}_1}&\Re\{\mathbf{b}_1^H\mathbf{G}^H\mathbf{G}\mathbf{b}_2\}\\
		\Re\{\mathbf{b}_2^H\mathbf{G}^H\mathbf{G}\mathbf{b}_1\}&{\mathbf{b}_2^H\mathbf{G}^H\mathbf{G}\mathbf{b}_2}
	\end{bmatrix}
\end{align}
The detailed derivation is provided in Appendix A. Therefore, the CRLBs corresponding to $\theta_{t,e}$ and $\phi_{t,e}$ can be respectively expressed as 
\begin{align}
	f_{\mathrm{CRLB}}(\theta_{t,e})&\geq[\boldsymbol{F}^{-1}]_{1,1}=\frac{\mathbf{b}_2^H\mathbf{G}^H\mathbf{G}\mathbf{b}_2}{2\hat{\sigma}_t^{-2} \det(\mathbf{F})}\\
	f_{\mathrm{CRLB}}(\phi_{t,e})&\geq[\boldsymbol{F}^{-1}]_{2,2}=\frac{\mathbf{b}_1^H\mathbf{G}^H\mathbf{G}\mathbf{b}_1}{2\hat{\sigma}_t^{-2} \det(\mathbf{F})}
\end{align}	
where
\begin{align}
	\det(\mathbf{F}) = &\ 2\hat{\sigma}_t^{-4} ( (\mathbf{b}_1^H\mathbf{G}^H\mathbf{G}\mathbf{b}_1)(\mathbf{b}_2^H\mathbf{G}^H\mathbf{G}\mathbf{b}_2) \notag\\&- |\Re\{\mathbf{b}_1^H\mathbf{G}^H\mathbf{G}\mathbf{b}_2\}|^2 )
\end{align}
	
	\subsubsection{Stage 2: SR maximization}
	According to \eqref{eq14} and \eqref{eq15}, the SINR corresponding to Bob and Eve can be denoted as
	\begin{align}
		\mathrm{SINR}_{t,b}&=\frac{|(\mathbf{h}_{t,Rb}^T\mathbf{\Theta}_t^c\mathbf{H}_t+\mathbf{h}_{ab}^T)\mathbf{w}_{t}|^2}{\sum_{l=1}^{L}|\mathbf{h}_{t,b,l}^T\mathbf{w}_t|^2+\sum_{l=1}^{L}|\mathbf{h}_{t,b,l}^T\mathbf{v}_t|^2+J_b}\\	\mathrm{SINR}_{t,e}&=\frac{|(\mathbf{h}_{t,Re}^T\mathbf{\Theta}_t^c\mathbf{H}_t+\mathbf{h}_{t,ae}^T)\mathbf{w}_{t}|^2}{\sum_{l=1}^{L}|\mathbf{h}_{t,e,l}^T\mathbf{w}_t|^2+\sum_{l=1}^{L}|\mathbf{h}_{t,e,l}^T\mathbf{v}_t|^2+J_e}
	\end{align}
	where \begin{align}J_b&=|(\mathbf{h}_{t,Rb}^T\mathbf{\Theta}_t^c\mathbf{H}_t+\mathbf{h}_{ab}^T)\mathbf{v}_{t}|^2+\sigma_{I}^{2}\|\mathbf{h}_{t,Rb}^T\mathbf{\Theta}_t^c\|+\sigma_t^2\\J_e&=|(\mathbf{h}_{t,Re}^T\mathbf{\Theta}_t^c\mathbf{H}_t+\mathbf{h}_{t,ae}^T)\mathbf{v}_{t}|^2+\sigma_{I}^{2}\|\mathbf{h}_{t,Re}^T\mathbf{\Theta}_t^c\|+\sigma_t^2\end{align} Then, the achievable transmission rates corresponding to Bob and Eve can be denoted as
	\begin{align}
		R_{t,b}&=\log_2(1+\mathrm{SINR}_{t,b})\\	R_{t,e}&=\log_2(1+\mathrm{SINR}_{t,e})
	\end{align}
	Correspondingly, the SR can be denoted as
	\begin{align}\label{36}
		R_{t,s} = [R_{t,b}-R_{t,e}]^+
	\end{align}
	
	The objective is to use the RIS orientation from stage 1 as the initial state and jointly optimize the RIS normal vector $\tilde{\mathbf{k}}_t$, the precoding vectors $\mathbf{w}_t$, and the phase shift matrices $\mathbf{\Theta}_t^c$ to maximize the SR. To ensure both the accuracy of subsequent DOA estimation and acceptable latency, the rotational range of the RIS is constrained to a limited interval. Consequently, the SR maximization problem can be formulated as
	\begin{subequations}
		\begin{align}
			\label{eq24a}
			\text{P2:}\quad&\quad\underset{\mathbf{w}_t,\mathbf{v}_t,\mathbf{\Theta}_t^c,\tilde{\mathbf{k}}_t}{\max}\ R_{t,s}
			\\ \label{eq24b}
			\mathrm{s.t.}\  &\text{C1}:\mathbf{w}_{t}^H\mathbf{w}_{t} +\mathbf{v}_{t}^H\mathbf{v}_{t}= P_t, \forall t\\\label{eq24c}
			&\text{C2}:\frac{\mathbf{b}_2^H\mathbf{G}^H\mathbf{G}\mathbf{b}_2}{2\hat{\sigma}_t^{-2} \det(\mathbf{F})}\leq\epsilon\\
			&\text{C3}:\frac{\mathbf{b}_1^H\mathbf{G}^H\mathbf{G}\mathbf{b}_1}{2\hat{\sigma}_t^{-2} \det(\mathbf{F})}\leq\epsilon\\
			&\text{C4}:\alpha_{\mathrm{min}}\leq\alpha_t\leq\alpha_{\max},\beta_{\mathrm{min}}\leq\beta_t\leq\beta_{\max},\forall t\\
			&\text{C5}:\varrho_{t,m}\leq \varrho_{\max},\forall t,m
		\end{align}
	\end{subequations}
	where $P_t$, $\epsilon$, and $\varrho_{\max}$ denote the transmission power budget, the CRLB threshold, and the maximum amplification factor of the RIS unit, respectively. Constraint C1 ensures that the total transmission power for the legitimate symbols and the AN symbols does not exceed the budget. Constraints C2 and C3 guarantee effective DOA estimation of Eve. Constraint C3 captures the achievable rotation range of the RIS. P2 is a non-convex optimization problem and is difficult to solve directly.

	\section{Proposed optimization algorithms}\label{sec:3}
In this section, three algorithms are proposed. Firstly, a Multi-stream Null-space Projection and Leakage-based (MNPL) method is developed. By employing null-space projection to design the beamforming at both the BS and the RIS, and leveraging leakage theory to mitigate interference. Secondly, an Enhanced Leakage-based (EL) method is proposed. This approach jointly considers both the LoS path and the RIS-reflected path to maximize the SR. The last one is a learning-based optimization scheme.
	\subsection{Proposed MNPL method}
	To maximize the received SINR during the DOA estimation phase, it is considered to align the RIS towards the position of Eve from the previous time slot, i.e., $(t-1)$-th slot. Note that due to Eve’s slow movement, the angle between Eve’s direction and the RIS normal is small. Therefore, given $\tilde{\mathbf{k}}_t$, P2 can be formulated as
	\begin{subequations}\label{38}
		\begin{align}
			\text{P3:}\quad&\underset{\mathbf{w}_t,\mathbf{v}_t,\mathbf{\Theta}_t^c}{\max}\ R_{t,s}
			\\ 
			\mathrm{s.t.}\  &\text{C1},\text{C5}
		\end{align}
	\end{subequations}
	Given the channels of Bob and Eve, the null-space projection matrix corresponding to BS can be constructed as
	\begin{align}\label{39}
		\mathbf{P}_{t,\mathrm{BS}}=\mathbf{I}_N-\mathbf{h}_{t,ae}^*(\mathbf{h}_{t,ae}^T\mathbf{h}_{t,ae}^*)^{-1}\mathbf{h}_{t,ae}^T
	\end{align}
	By introducing the auxiliary variable $\bm{\zeta}_{t,\mathrm{BS}}$, we have
	\begin{align}\label{40}
		\mathbf{w}_t=\mathbf{P}_{t,\mathrm{BS}}\bm{\zeta}_{t,\mathrm{BS}}
	\end{align}
	Similarly, the null-space projection matrix corresponding to the RIS can be denoted as
	\begin{align}\label{41}
		\mathbf{P}_{t,\mathrm{RIS}}=\mathbf{I}_M-\bm{f}_t(\bm{f}_t^H\bm{f}_t)^{-1}\bm{f}_t^H
	\end{align}
	where
	\begin{align}
		\bm{f}_t=\mathrm{diag}(\mathbf{h}_{t,Re})\mathbf{H}_t\mathbf{P}_{t,\mathrm{BS}}\bm{\zeta}_{t,\mathrm{BS}}
	\end{align}
	By introducing the auxiliary variable $\bm{\zeta}_{t,\mathrm{RIS}}$, we have
	\begin{align}\label{43}
		\bm{\upsilon}_t^c=\mathbf{P}_{t,\mathrm{RIS}}^T\bm{\zeta}_{t,\mathrm{RIS}}.
	\end{align}
	Since the Eve is unable to receive the confidential signal, the entire transmit power and RIS reflection power are allocated for the confidential signal transmission. Therefore, by substituting \eqref{40} and \eqref{43} into \eqref{38}, P3 can be reformulated as
	\begin{subequations}
		\begin{align}
			\text{P4:}\quad&\underset{\bm{\zeta}_{t,\mathrm{BS}},\bm{\zeta}_{t,\mathrm{RIS}}}{\max}\ \frac{\bm{\zeta}_{t,\mathrm{RIS}}^H\mathbf{A}_1\bm{\zeta}_{t,\mathrm{RIS}}+\bm{\zeta}_{t,\mathrm{BS}}^H\mathbf{A}_2\bm{\zeta}_{t,\mathrm{BS}}}{\bm{\zeta}_{t,\mathrm{BS}}^H\mathbf{A}_3\bm{\zeta}_{t,\mathrm{BS}}+\bm{\zeta}_{t,\mathrm{RIS}}^H\mathbf{A}_4\bm{\zeta}_{t,\mathrm{RIS}}+\sigma_t^2}
			\\ 
			\mathrm{s.t.}\  &\text{C1}^{\prime}:\bm{\zeta}_{t,\mathrm{BS}}^H\mathbf{P}_{t,\mathrm{BS}}^H\mathbf{P}_{t,\mathrm{BS}}\bm{\zeta}_{t,\mathrm{BS}} = P_t, \forall t\\
			&\text{C5}
		\end{align}
	\end{subequations}
	where
	\begin{align}
		\mathbf{A}_1 &= \mathbf{P}_{\mathrm{RIS}}\bm{f}_t \bm{f}_t^H\mathbf{P}_{\mathrm{RIS}}^H\\
		\mathbf{A}_2 &=\mathbf{P}_{t,\mathrm{BS}}^H\mathbf{h}_{ab}^*\mathbf{h}_{ab}^T\mathbf{P}_{t,\mathrm{BS}}\\
		\mathbf{A}_3 &= \sum_{l=1}^{L}\mathbf{P}_{t,\mathrm{BS}}^H\mathbf{h}_{t,b,l}^*\mathbf{h}_{t,b,l}^T\mathbf{P}_{t,\mathrm{BS}}\\
		\mathbf{A}_4 &= \sigma_{I}^{2}\mathbf{P}_{\mathrm{RIS}}\mathrm{diag}(\mathbf{h}_{t,Rb})\mathrm{diag}(\mathbf{h}_{t,Rb}^*)\mathbf{P}_{\mathrm{RIS}}^H
	\end{align}
	Due to the coupling between variables $\bm{\zeta}_{t,\mathrm{BS}}$ and $\bm{\zeta}_{t,\mathrm{RIS}}$, P4 remains a non-convex optimization problem. To tackle this challenge, a two-step solution is proposed. In practice, considering that the RIS-reflected path may suffer from multiplicative path loss and amplified noise interference, its channel condition can be weaker than that of the LoS path. Therefore, the precoding vector $\mathbf{w}_t$ is optimized first, followed by the optimization of the phase-shift matrix $\bm{\Theta}_t^c$. Specifically, the sub-optimization problem with respect to $\bm{\zeta}_{t,\mathrm{BS}}$ can be formulated as
	\begin{subequations}\label{49}
		\begin{align}
			\text{P5:}\quad&\underset{\bm{\zeta}_{t,\mathrm{BS}}}{\max}\ \frac{\bm{\zeta}_{t,\mathrm{BS}}^H\mathbf{A}_2\bm{\zeta}_{t,\mathrm{BS}}}{\bm{\zeta}_{t,\mathrm{BS}}^H\mathbf{A}_3\bm{\zeta}_{t,\mathrm{BS}}}
			\\ 
			\mathrm{s.t.}\  &\text{C1}^{\prime}
		\end{align}
	\end{subequations}
	P5 is a generalized Rayleigh quotient problem. The solution for $\bm{\zeta}_{t,\mathrm{BS}}$ is the eigenvector $\bm{\tau}_{t,\mathrm{BS}}$ corresponding to the largest eigenvalue of $\mathbf{A}_3^{-1}\mathbf{A}_2$, and we have
	\begin{align}\label{50}
		\mathbf{w}_t=\sqrt{P_t}\frac{\mathbf{P}_{t,\mathrm{BS}}\bm{\tau}_{t,\mathrm{BS}}}{\|\mathbf{P}_{t,\mathrm{BS}}\bm{\tau}_{t,\mathrm{BS}}\|}
	\end{align}
	Given $\mathbf{w}_t$, the subproblem with respect to $\bm{\zeta}_{t,\mathrm{RIS}}$ can be formulated as
	\begin{subequations}
		\begin{align}
			\text{P5:}\quad&\underset{\bm{\zeta}_{t,\mathrm{RIS}}}{\max}\ \frac{\bm{\zeta}_{t,\mathrm{RIS}}^H\mathbf{A}_1\bm{\zeta}_{t,\mathrm{RIS}}}{\bm{\zeta}_{t,\mathrm{RIS}}^H\mathbf{A}_4\bm{\zeta}_{t,\mathrm{RIS}}}
			\\ 
			\mathrm{s.t.}\  &\text{C5}.
		\end{align}
	\end{subequations}
	By relaxing C5, the solution for $\bm{\zeta}_{t,\mathrm{RIS}}$ can be obtained as 
	\begin{align}\label{52}
		\bm{\upsilon}_t^c=\frac{\varrho_{\max}\bm{\tau}_{t,\mathrm{RIS}}^T\mathbf{P}_{t,\mathrm{RIS}}}{\|\bm{\tau}_{t,\mathrm{RIS}}^T\mathbf{P}_{t,\mathrm{RIS}}\|_{\infty}}
	\end{align}
	where $\bm{\tau}_{t,\mathrm{RIS}}$ denotes the eigenvector corresponding to the largest eigenvalue of $\mathbf{A}_4^{-1}\mathbf{A}_1$. For clarity, the details are shown in Algorithm \ref{alg1}.
	\begin {algorithm}[h] \footnotesize
	\caption{The proposed MNPL method}\label{alg1}
	\textbf{Input:} $P_t$, $\varrho_{\max}$, $\{\bm{\Theta}_t^p\}_{p=1}^{P}$, $N$, $M$, $L$, $\sigma_I^2$, $\tilde{\mathbf{k}}_t$\\
	\textbf{Output:} $\mathbf{w}_t$, $\mathbf{\Theta}_t^c$
	
	\begin{algorithmic}[1]
		\State Compute the null-space projection matrix using \eqref{39} and \eqref{41}.
		\State For the LoS and multipath interference components, utilize \eqref{50} to maximize the received SINR at Bob.
		\State Given $\mathbf{w}_t$, compute $\bm{\upsilon}_t^c$ using \eqref{52} to maximize the received SINR corresponding to the RIS-reflected path.
		\State$\mathbf{w}_t$, $\mathbf{\Theta}_t^c$.	
	\end{algorithmic}
	\end{algorithm}
	\subsection{Proposed EL method}
	To maximize the SR, according to the leakage theory, the energy corresponding to confidential information transmission is suppressed in the directions of Eve and the distinguishable equivalent multipath interference directions. Meanwhile, the focusing effect of AN toward Bob is reduced. Specifically, P2 can be reformulated as
	\begin{subequations}
		\begin{align}
			&\text{P6:}\quad\underset{\mathbf{w}_t,\mathbf{v}_t,\bm{\upsilon}_t^c}{\max}\ O_1=\notag\\& \frac{\mathbf{w}_t^H\mathbf{A}_5\mathbf{w}_t+\mathbf{v}_t^H(\mathbf{A}_6+2\mathbf{A}_8)\mathbf{v}_t+\sigma_{I}^{2}(\bm{\upsilon}_t^c)^H\mathbf{A}_{9}\bm{\upsilon}_t^c}{\mathbf{w}_t^H(\mathbf{A}_6+\mathbf{A}_7)\mathbf{w}_t+\mathbf{v}_t^H(\mathbf{A}_5+\mathbf{A}_7)\mathbf{v}_t+\sigma_{I}^{2}(\bm{\upsilon}_t^c)^H\mathbf{A}_{10}\bm{\upsilon}_t^c}
			\\ 
			&\mathrm{s.t.}\  \text{C1},\text{C5}
		\end{align}
	\end{subequations}
	where
	\begin{align}
		\mathbf{A}_5 &= (\mathbf{h}_{t,Rb}^T\mathbf{\Theta}_t^c\mathbf{H}_t+\mathbf{h}_{ab}^T)^H(\mathbf{h}_{t,Rb}^T\mathbf{\Theta}_t^c\mathbf{H}_t+\mathbf{h}_{ab}^T)\\
		\mathbf{A}_6 &=(\mathbf{h}_{t,Re}^T\mathbf{\Theta}_t^c\mathbf{H}_t+\mathbf{h}_{ae}^T)^H(\mathbf{h}_{t,Re}^T\mathbf{\Theta}_t^c\mathbf{H}_t+\mathbf{h}_{ae}^T)-\mathbf{A}_8\\
		\mathbf{A}_7 &= \sum_{l=1}^{L}\mathbf{h}_{t,b,l}^*\mathbf{h}_{t,b,l}^T\\
		\mathbf{A}_8 &= \sum_{l=1}^{L}\mathbf{h}_{t,e,l}^*\mathbf{h}_{t,e,l}^T\\
		\mathbf{A}_{9} &= \mathbf{h}_{t,Re}\mathbf{h}_{t,Re}^H\\
		\mathbf{A}_{10} &= \mathbf{h}_{t,Rb}\mathbf{h}_{t,Rb}^H\\
		\mathbf{A}_{11} &= \mathrm{diag}(\mathbf{h}_{t,Re})\mathrm{diag}(\mathbf{h}_{t,Re}^*)\\
		\mathbf{A}_{12} &= \mathrm{diag}(\mathbf{h}_{t,Rb})\mathrm{diag}(\mathbf{h}_{t,Rb}^*)
	\end{align}
	Here, $\mathbf{A}_5$ denotes the channel covariance matrix corresponding to Bob, while $\mathbf{A}_6$ represents the channel covariance matrix corresponding to Eve, with the inclusion of multipath interference at Eve. $\mathbf{A}_7$ and $\mathbf{A}_8$ are the covariance matrices associated with the multipath interference for Bob and Eve, respectively. $\mathbf{A}_9$ and $\mathbf{A}_{10}$ denote the covariance matrices of the channels from the RIS to Eve and Bob, respectively.
	
	Due to the highly nonlinear coupling among multiple variables, P6 is difficult to solve directly. Therefore, a stepwise optimization approach is employed. Specifically, with $\mathbf{w}_t$ and $\mathbf{v}_t$ fixed, the subproblem with respect to $\bm{\upsilon}_t^c$ can be formulated as
	\begin{subequations}
		\begin{align}
			&\text{P7:}\quad\underset{\bm{\upsilon}_t^c}{\max}\  \frac{(\bm{\upsilon}_t^c)^H(\mathbf{A}_{9}+\sigma_{I}^{2}\mathbf{A}_{11}+\mathbf{A}_{10})\bm{\upsilon}_t^c}{\sigma_{I}^{2}(\bm{\upsilon}_t^c)^H\mathbf{A}_{12}\bm{\upsilon}_t^c}
			\\ 
			&\mathrm{s.t.}\  \text{C5}
		\end{align}
	\end{subequations}
	P7 aims to maximize the energy transmitted from RIS to both Bob and Eve, while simultaneously reducing the RIS-amplified noise power at Bob and increasing it at Eve. According to the leakage theory, the solution for $\bm{\upsilon}_t^c$ is the eigenvector $\hat{\tau}_{\mathrm{RIS}}$ corresponding to the largest eigenvalue of $(\sigma_{I}^{2}\mathbf{A}_{12})^{-1}((1+\sigma_{I}^{2})\mathbf{A}_{11}+\mathbf{A}_{12})$. After normalization, $\bm{\upsilon}_t^c$ can be expressed as
	\begin{align}\label{62}
		\bm{\upsilon}_t^c = \frac{\varrho_{\max}\hat{\bm{\tau}}_{t,\mathrm{RIS}}}{\|\hat{\bm{\tau}}_{t,\mathrm{RIS}}\|_{\infty}}
	\end{align}
	
	According to P6, given $\mathbf{w}_t$ and $\bm{\upsilon}_t^c$, the subproblem with respect to $\mathbf{v}_t$ can be formulated as
	\begin{subequations}\label{63}
		\begin{align}
			&\text{P8:}\quad\underset{\mathbf{v}_t}{\max}\  \frac{\mathbf{v}_t^H(\mathbf{A}_6+2\mathbf{A}_8)\mathbf{v}_t+A_1}{\mathbf{v}_t^H(\mathbf{A}_5+\mathbf{A}_7)\mathbf{v}_t+A_2}
			\\ 
			&\mathrm{s.t.}\  \text{C6}:\mathbf{v}_t^H\mathbf{v}_t=1
		\end{align}
	\end{subequations}
	where $A_1 = \mathbf{w}_t^H\mathbf{A}_5\mathbf{w}_t+\sigma_{I}^{2}(\bm{\upsilon}_t^c)^H\mathbf{A}_{11}\bm{\upsilon}_t^c$ and $A_2 = \mathbf{w}_t^H(\mathbf{A}_6+\mathbf{A}_7)\mathbf{w}_t+\sigma_{I}^{2}(\bm{\upsilon}_t^c)^H\mathbf{A}_{12}\bm{\upsilon}_t^c$. P8 maximizes the SR by designing AN with unit power. The solution for $\mathbf{v}_t$ is the eigenvector $\hat{\bm{\tau}}_{\mathrm{BS},v_t}$ corresponding to the largest eigenvalue of $(\mathbf{A}_5+\mathbf{A}_7+A_2\mathbf{I}_N)^{-1}(\mathbf{A}_6+2\mathbf{A}_8+A_1\mathbf{I}_N)$.
	
	Correspondingly, given $\mathbf{v}_t$ and $\bm{\upsilon}_t^c$, the subproblem with respect to $\mathbf{w}_t$ can be formulated as
	\begin{subequations}\label{64}
		\begin{align}
			&\text{P9:}\quad\underset{\mathbf{w}_t}{\max}\  \frac{\mathbf{w}_t^H\mathbf{A}_5\mathbf{w}_t+A_3}{\mathbf{w}_t^H(\mathbf{A}_6+\mathbf{A}_7)\mathbf{w}_t+A_4}
			\\ 
			&\mathrm{s.t.}\  \text{C6}:\mathbf{w}_t^H\mathbf{w}_t=1
		\end{align}
	\end{subequations}
	where $A_3 = \mathbf{v}_t^H(\mathbf{A}_6+2\mathbf{A}_8)\mathbf{v}_t+\sigma_{I}^{2}(\bm{\upsilon}_t^c)^H\mathbf{A}_{11}\bm{\upsilon}_t^c$ and $A_4 = \mathbf{v}_t^H(\mathbf{A}_5+\mathbf{A}_7)\mathbf{v}_t+\sigma_{I}^{2}(\bm{\upsilon}_t^c)^H\mathbf{A}_{12}\bm{\upsilon}_t^c$. The solution for $\mathbf{w}_t$ is the eigenvector $\hat{\bm{\tau}}_{\mathrm{BS},w_t}$ corresponding to the largest eigenvalue of $(\mathbf{A}_6+\mathbf{A}_7+A_4\mathbf{I}_N)^{-1}(\mathbf{A}_5+A_3\mathbf{I}_N)$.
	
	To satisfy the power constraint, an auxiliary variable $\xi_t$ is introduced to further improve the SR. Substituting \eqref{62}, $\hat{\bm{\tau}}_{\mathrm{BS},w_t}$, and $\hat{\bm{\tau}}_{\mathrm{BS},v_t}$ into \eqref{36}, we can obtain \eqref{69}, where
	\begin{figure*}
		\begin{align}\label{69}
			\text{P10:}\quad\underset{\xi_t}{\max}\ \log_2\left( \frac{1+\frac{\xi_tP_t\hat{\bm{\tau}}_{\mathrm{BS},w_t}^H\mathbf{H}_{t,\mathrm{all},b}^H\mathbf{H}_{t,\mathrm{all},b}\hat{\bm{\tau}}_{\mathrm{BS},w_t}}{\xi_tP_t\hat{\bm{\tau}}_{\mathrm{BS},w_t}^H\mathbf{A}_7\hat{\bm{\tau}}_{\mathrm{BS},w_t}+(1-\xi_t)P_t\hat{\bm{\tau}}_{\mathrm{BS},v_t}^H(\mathbf{A}_7+\mathbf{H}_{t,\mathrm{all},b}^H\mathbf{H}_{t,\mathrm{all},b})\hat{\bm{\tau}}_{\mathrm{BS},v_t}+A_5}}{1+\frac{\xi_tP_t\hat{\bm{\tau}}_{\mathrm{BS},w_t}^H\mathbf{H}_{t,\mathrm{all},e}^H\mathbf{H}_{t,\mathrm{all},e}\hat{\bm{\tau}}_{\mathrm{BS},w_t}}{\xi_tP_t\hat{\bm{\tau}}_{\mathrm{BS},w_t}^H\mathbf{A}_8\hat{\bm{\tau}}_{\mathrm{BS},w_t}+(1-\xi_t)P_t\hat{\bm{\tau}}_{\mathrm{BS},v_t}^H(\mathbf{A}_8+\mathbf{H}_{t,\mathrm{all},e}^H\mathbf{H}_{t,\mathrm{all},e})\hat{\bm{\tau}}_{\mathrm{BS},v_t}+A_5}}\right) \quad \mathrm{s.t.}\ 0\leq\xi_t\leq1
		\end{align}
	\end{figure*}
	
	\begin{align} &\mathbf{H}_{t,\mathrm{all},b}=\mathbf{h}_{t,Rb}^T\mathrm{diag}(\frac{\varrho_{\max}\hat{\bm{\tau}}_{t,\mathrm{RIS}}}{\|\hat{\bm{\tau}}_{t,\mathrm{RIS}}\|_{\infty}})\mathbf{H}_t+\mathbf{h}_{ab}^T\\ &\mathbf{H}_{t,\mathrm{all},e}=\mathbf{h}_{t,Re}^T\mathrm{diag}(\frac{\varrho_{\max}\hat{\bm{\tau}}_{t,\mathrm{RIS}}}{\|\hat{\bm{\tau}}_{t,\mathrm{RIS}}\|_{\infty}})\mathbf{H}_t+\mathbf{h}_{ae}^T\\ &A_5=\sigma_{I}^{2}\|\mathbf{h}_{t,Rb}^T\mathrm{diag}(\frac{\varrho_{\max}\hat{\bm{\tau}}_{t,\mathrm{RIS}}}{\|\hat{\bm{\tau}}_{t,\mathrm{RIS}}\|_{\infty}})\|+\sigma_t^2\\ &A_6=\sigma_{I}^{2}\|\mathbf{h}_{t,Re}^T\mathrm{diag}(\frac{\varrho_{\max}\hat{\bm{\tau}}_{t,\mathrm{RIS}}}{\|\hat{\bm{\tau}}_{t,\mathrm{RIS}}\|_{\infty}})\|+\sigma_t^2
	\end{align}
	P10 is a fractional programming problem, which can be solved using the classical Dinkelbach algorithm or a one-dimensional search to obtain the optimal $\xi_t$. Then, $\mathbf{w}_t$ and $\mathbf{v}_t$ can be expressed as $\mathbf{w}_t=\sqrt{\xi_tP_t}\hat{\bm{\tau}}_{\mathrm{BS},w_t}$ and $\mathbf{v}_t=\sqrt{(1-\xi_t)P_t}\hat{\bm{\tau}}_{\mathrm{BS},v_t}$, respectively. For clarity, the details are shown in Algorithm \ref{alg2}.
		\begin {algorithm}[h] \footnotesize
	\caption{The proposed EL method}\label{alg2}
	\textbf{Input:} $P_t$, $\varrho_{\max}$, $\{\bm{\Theta}_t^p\}_{p=1}^{P}$, $N$, $M$, $L$, $\sigma_I^2$, $\tilde{\mathbf{k}}_t$\\
	\textbf{Output:} $\mathbf{w}_t$, $\mathbf{v}_t$, $\mathbf{\Theta}_t^c$
	
	\begin{algorithmic}[1]
			\State Initialize $\mathbf{w}_t$.
\State Compute $\bm{\upsilon}_t^c$ based on \eqref{62}.
\State Given $\mathbf{w}_t$ and $\bm{\upsilon}_t^c$, compute $\hat{\bm{\tau}}_{\mathrm{BS},w_t}$ using the leakage theory according to \eqref{63}.
\State Given $\hat{\bm{\tau}}_{\mathrm{BS},v_t}$ and $\bm{\upsilon}_t^c$, compute $\mathbf{w}_t$ using the leakage theory according to \eqref{64}.
\State According to \eqref{69}, obtain $\xi_t$ using the Dinkelbach algorithm or a one-dimensional search.
\State Given $\xi_t$, compute $\mathbf{w}_t$ and $\mathbf{v}_t$.
\State\textbf{return} $\mathbf{w}_t$, $\mathbf{v}_t$, $\mathbf{\Theta}_t^c$, and $\xi_t$.	
	\end{algorithmic}
	\end{algorithm}
	\subsection{Proposed Optimization Scheme Based on DSAC-T}
	Algorithm \ref{alg1} and Algorithm \ref{alg2} achieve secure beamforming design under guaranteed DOA estimation accuracy by aligning the RIS with the Eve's location. However, there exists a trade-off between security performance and DOA estimation accuracy. Conventional iterative optimization algorithms typically require a large number of iterations and are prone to getting stuck in local optima. Through exploration and learning, reinforcement learning holds the potential to achieve higher security performance offline. The overestimation of $Q$-values leading to suboptimal policies is a significant challenge in reinforcement learning algorithms. Compared to the distributed soft actor-critic (DSAC) algorithm~\cite{Duan2022}, an enhanced version termed DSAC with Three refinements (DSAC-T) has been validated to ensure the stability of the learning process and maintain robust performance across different reward scales~\cite{Duan2025}. The DSAC-T algorithm demonstrates significant advantages in handling continuous action spaces and high-dimensional state spaces. Moreover, its introduced dual-softening mechanism effectively balances exploration and exploitation, thereby avoiding entrapment in locally optimal solutions. Therefore, an optimization scheme based on DSAC-T is presented in this section. Specifically, we define the following mapping relationship.
	\begin{enumerate}
		\item  State space: We consider the physical state of wireless communications and the associated system state as the state space, which can be represented as $\mathcal{S}$. Then, the state $s^{z}\in\mathcal{S}$ of $z$-th step can be denoted as
		\begin{equation}
			s^{z}=\{b_1^{z},b_2^{z},[\boldsymbol{F}^{-1}]_{1,1}^{z},[\boldsymbol{F}^{-1}]_{2,2}^{z}\}
		\end{equation}
		where $b_1=\mathbf{w}_{t}^H\mathbf{w}_{t}$ and $b_2=\mathbf{v}_{t}^H\mathbf{v}_{t}$.
		\item   Action space:  The RIS acts as an agent capable of rotation, thereby influencing both the DOA estimation accuracy and security performance. Therefore, the action $a^{z}$ can be represented as
		\begin{equation}
			a^{z}=\{\tilde{\mathbf{k}}_t^{z}\}\in\mathcal{A}
		\end{equation}
		where $\mathcal{A}$ denotes the action space.
		\item   Reward:  We design the reward as follows:
		\begin{equation}
			r^{z}=R_{t,s}^{z}-\Gamma
		\end{equation}
		where $\Gamma$ denotes the penalty factor. Due to the framework where DOA estimation is performed first, followed by beamforming design, it can be considered that the agent's action primarily impacts the DOA estimation performance. Therefore, when constraints C2 and C3 are not satisfied, the reward is designed to be zero, indicating that security performance is not guaranteed.
		\item  Environment: The underlying idea is to leverage reinforcement learning to improve upon traditional convex optimization algorithms, thereby enhancing system performance. Consequently, Algorithm 1 and Algorithm 2 are considered part of the environment, where the actions taken by the agent are fed into these algorithms to interact with the physical environment and generate new states.
		
	\end{enumerate}
	
	Based on DSAC-T, we incorporate policy entropy into the reward signal, leading to an entropy-augmented objective function, denoted as 
	\begin{equation}\begin{aligned}&J_{\pi}=\underset{(s^{i\geq z},a^{i\geq z})\sim\rho_{\pi}}{\mathbb{E}}\left[\sum_{i=z}^{\infty}\gamma^{i-z}[r^{i}+\iota\mathcal{H}(\pi(\cdot|s^{i}))]\right]\end{aligned}\end{equation}
	where $\rho_{\pi}$ represents the distribution of trajectories induced by policy $\pi$ and $\mathcal{H}(\pi(\cdot|s))=\mathbb{E}_{a\sim\pi(\cdot|s)}[-\log\pi(a|s)]$. $\gamma$ and $\iota$ denote the discount factor and the temperature coefficient, respectively. The $Q$-value can be denoted as
	\begin{align}&Q^{\pi}(s^{z},a^{z})\notag\\&=r^{z}+\gamma\underset{(s^{i>z},a^{i>z})\sim\rho_{\pi}}{\mathbb{E}}[\sum_{i=z+1}^{\infty}\gamma^{i-z-1}[r^{i}-\iota\log\pi(a^{i}|s^{i})]]\end{align}
	Then, soft policy evaluation and soft policy improvement are carried out. Specifically, the $Q$-value is expressed via the Bellman equation as the sum of the current reward and the discounted future $Q$‑values for critic update. Alternately, based on the current $Q$‑value, the policy is updated to maximize the sum of $Q$‑value and entropy, thereby achieving actor update.
	\subsubsection{ Critic update}
	For DSAC-T, both the value distribution and the stochastic policy can be modeled as diagonal Gaussians, parameterized as \begin{equation} \mathcal{Z}_{\theta}(\cdot \mid s, a)=\mathcal{N}(Q_\theta(s,a),\sigma_\theta(s,a)^2) \end{equation}and \( \pi_{\phi}(\cdot \mid s) \), respectively. To reduce the high randomness arising from the \( \mathcal{Z}_{\theta}(\cdot \mid s, a) \) in the target return, we consider replacing the target return $y_1=r+\iota(\mathcal{Z}(s^{\prime},a^{\prime})-\alpha\log\pi_{\bar{\phi}}(a^{\prime}|s^{\prime}))$ with the target $Q$-value, leading to
	\begin{equation}y_2=r+\gamma(Q_{\bar{\theta}}(s^{\prime},a^{\prime})-\iota\log\pi_{\bar{\phi}}(a^{\prime}|s^{\prime}))\end{equation}
	To prevent gradient explosion caused by variance-related gradients, a clipping function 
	\begin{equation}C(y_1;\bar{b}):=\operatorname{clip}\left(y_1,Q_\theta(s,a)-\bar{b},Q_\theta(s,a)+\bar{b}\right)\end{equation}
	is introduced, where
	\begin{equation}\bar{b}=\hat{\xi}\underset{(s,a)\sim\mathcal{B}}{\operatorname*{\mathbb{E}}}[\sigma_{\theta}(s,a)]\end{equation}
	is the clipping boundary. $\mathcal{B}$ and $\hat{\xi}$ denote the replay buffer and a constant parameter controlling the clipping range, respectively. Then, the critic gradient is expressed as 
	\begin{equation}\begin{aligned}\nabla_{\theta}J_{\mathcal{Z}}(\theta)&\approx\mathbb{E}\Big[-\frac{(y_{2}-Q_{\theta}(s,a))}{\sigma_{\theta}(s,a)^{2}}\nabla_{\theta}Q_{\theta}(s,a)\\&-\frac{(C(y_{1};\bar{b})-Q_{\theta}(s,a))^{2}-\sigma_{\theta}(s,a)^{2}}{\sigma_{\theta}(s,a)^{3}}\nabla_{\theta}\sigma_{\theta}(s,a)\Big]\end{aligned}\end{equation}
	To prevent overestimation, twin value distribution learning is adopted, where the smaller values $y_{1}^{\min}$ and $y_{2}^{\min}$ are selected from the two networks for gradient updates. To mitigate the learning sensitivity to reward scale, a gradient scaling weight $\omega=\mathbb{E}_{(s,a)\sim\mathcal{B}}\begin{bmatrix}\sigma_\theta(s,a)^2\end{bmatrix}$ is introduced. Then, the corresponding gradient for each value distribution can be denoted as
	\begin{equation}\begin{aligned}\nabla_{\theta_{i}}J_{\mathcal{Z}}^{\mathrm{s}}(\theta_{i})\approx&(\omega_{i}+\epsilon_{\omega})\mathbb{E}\left[-\frac{\left(y_{2}^{\min}-Q_{\theta_{i}}(s,a)\right)}{\sigma_{\theta_{i}}(s,a)^{2}+\epsilon}\nabla_{\theta_{i}}Q_{\theta_{i}}(s,a)\right]\\&\left.-\frac{\left(C(y_{1}^{\min},\bar{b}_{i})-Q_{\theta_{i}}(s,a)\right)^{2}-\sigma_{\theta_{i}}(s,a)^{2}}{\sigma_{\theta_{i}}(s,a)^{3}+\epsilon}\nabla_{\theta_{i}}\sigma_{\theta_{i}}(s,a)\right]\end{aligned}\end{equation}
	where $\epsilon$ and $\epsilon_{\omega}$ are small positive constants introduced to prevent gradient explosion and vanishing gradients, respectively.
	\begin{equation}\begin{aligned}&\bar{b}_{i}\leftarrow\bar{\tau}\hat{\xi}\mathbb{E}_{(s,a)\sim\mathcal{B}}[\sigma_{\theta_{i}}(s,a)]+(1-\bar{\tau})\bar{b}_{i}\\&\omega_{i}\leftarrow\bar{\tau}\mathbb{E}_{(s,a)\sim\mathcal{B}}\left[\sigma_{\theta_{i}}(s,a)^{2}\right]+(1-\bar{\tau})\omega_{i}\end{aligned}\end{equation}
	where $\bar{\tau}$ denotes the synchronization rate.
	\subsubsection{ Actor update}
	By improving the policy to perform action updates, we obtain
	\begin{equation}J_{\pi}(\phi)=\underset{s\sim\mathcal{B},a\sim\pi_{\phi}}{\mathbb{E}}[Q_{\theta}(s,a)-\iota\log(\pi_{\phi}(a|s))]\end{equation}
	The temperature parameter $\iota$ is updated as
	\begin{equation}\iota\leftarrow\iota-\beta_\iota\underset{s\sim\mathcal{B},a\sim\pi_\phi}{\mathbb{E}}[-\log\pi_\phi(a|s)-\overline{\mathcal{H}}]\end{equation}
	where $\mathcal{H}$ and $\beta_\iota$ denote the target entropy and  the learning rate, respectively. Similarly, the actor's target undergoes correction based on the twin value distributions, which can be expressed as 
	\begin{equation}J_{\pi}(\phi)=\underset{s\sim\mathcal{B},a\sim\pi_{\phi}}{\mathbb{E}}[\min_{i=1,2}Q_{\theta_{i}}(s,a)-\iota\log(\pi_{\phi}(a|s))]\end{equation}
	
\begin{figure}
		\centering
		\includegraphics[width=0.45\textwidth, trim = 50 15 20 2,clip]{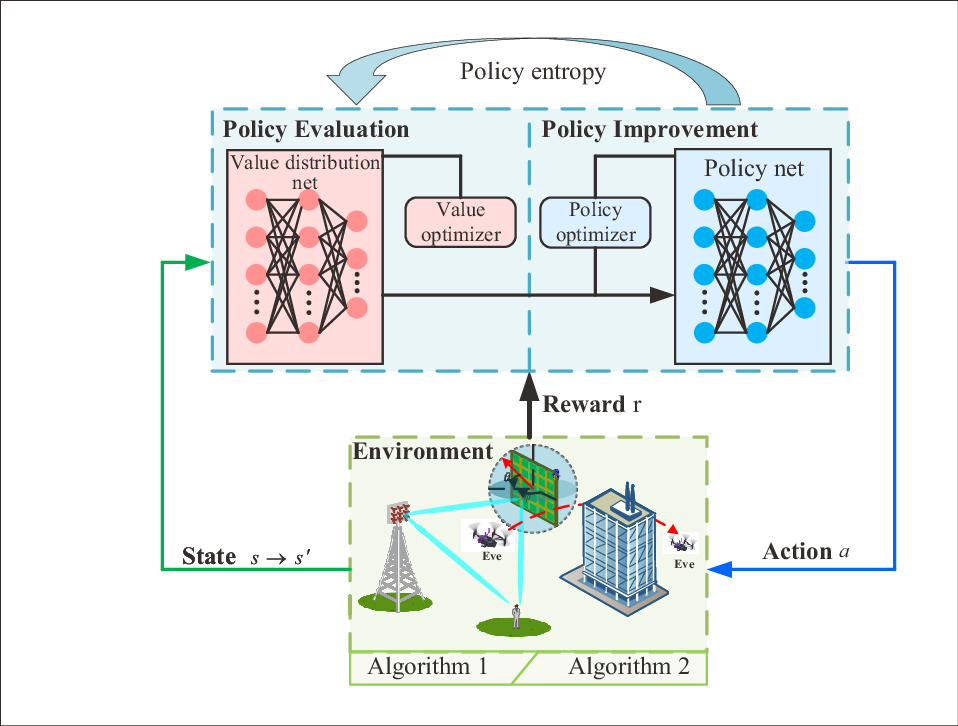}\\
		\caption{Optimization scheme based on DSAC-T.}\label{fig:2}
	\end{figure}

	For clarity, the optimization scheme based on DSAC-T is illustrated in Fig. \ref{fig:2}. For simplicity, this optimization scheme is referred to as Algorithm 3.

	\section{Simulation results}\label{sec:4}
	
	The simulation results are presented in this section. Firstly, the system parameters are provided. We consider two distinguishable data streams: data stream 1 corresponds to the LoS path, and data stream 2 corresponds to the RIS‑reflected path. Unless otherwise specified, the numbers of antennas at the BS and RIS elements are $N = 16$ and $M = 64$, respectively. The carrier frequency, transmit power, and maximum reflection coefficient of the RIS elements are set to $2.4$ GHz, $P_t = 40$ dBm, and $\varrho_{\mathrm{max}}=10$, respectively. The ranges of the elevation and azimuth angles associated with the $\tilde{\mathbf{k}}$ are denoted as $-80^{\circ}\leq\alpha_t\leq0^{\circ}$ and $90^{\circ}\leq\beta_t\leq150^{\circ}$, respectively. The distances from the RIS to the BS, Eve, and Bob are 20 m, 5 m, and 10 m, respectively. Assume that the RIS and Eve are at the same height, i.e., 10 m. The noise experienced by the RIS, Bob, and Eve is denoted as $-90$ dBm, $-80$ dBm, and $-80$ dBm, respectively. The hyperparameters corresponding to the DSAC-T-based optimization scheme are listed in Table \ref{lab1}.
	\begin{table}[t]\footnotesize
		\centering
		\begin{threeparttable}
			\setlength{\abovecaptionskip}{0pt}
			\caption{Detailed Hyperparameters.}\label{lab1}
			\begin{tabular}{cc}
				\hline
				Hyperparameters& Value \\
				\hline
				Actor learning rate & 3e-4 \\
				Critic learning rate & 3e-4\\
				Discount factor & 0.99\\
				Soft update coefficient&0.005\\
				Temperature parameter&0.01\\
				Value distribution range&(0,30)\\
				Hidden layer dimension&256\\
				Replay buffer size&100000\\
				Sample batch size&64\\
				Adaptive bound range&[0.5,2.0]\\
				\hline
			\end{tabular}
		\end{threeparttable}
	\end{table}
	
\begin{figure}[t]
	\centering
	\includegraphics[width=0.45\textwidth]{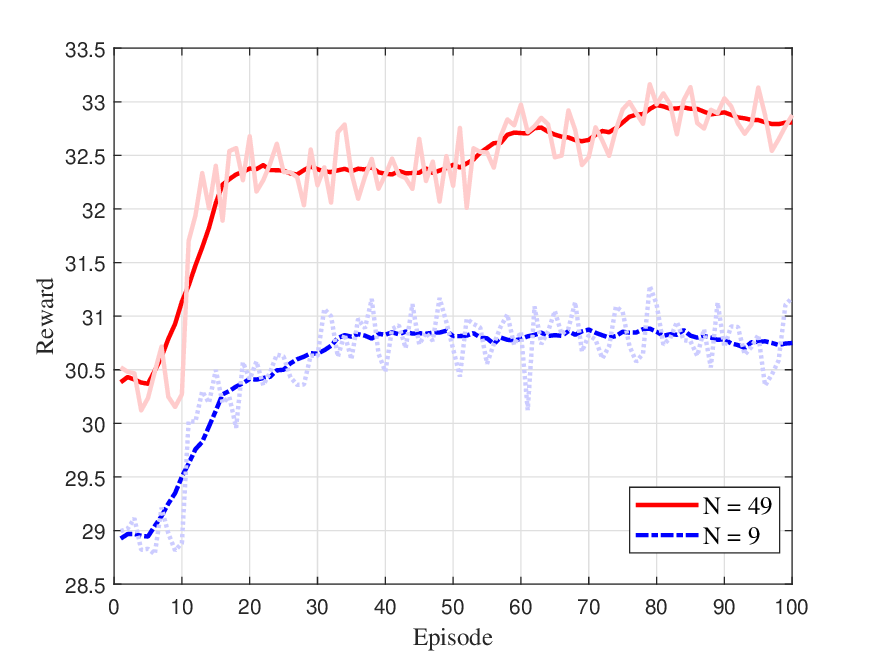}\\
	\caption{ Reward versus training episode.}\label{fig3}
\end{figure}

	The reward curve of Algorithm 3 is illustrated in Fig. \ref{fig3}. The solid red line represents \( N = 49 \), and the dashed blue line represents \( N = 9 \). The lighter curves indicate the raw reward fluctuations, while the darker curves are the results after smoothing with a sliding window. As the number of episodes increases, the reward values tend to stabilize. 
	\begin{figure}[htbp]
		\centering   
		\subfigure[] 
		{
			\begin{minipage}[b]{0.4\linewidth}
				\centering
				\includegraphics[width=1.2\textwidth, trim = 10 2 2 0,clip]{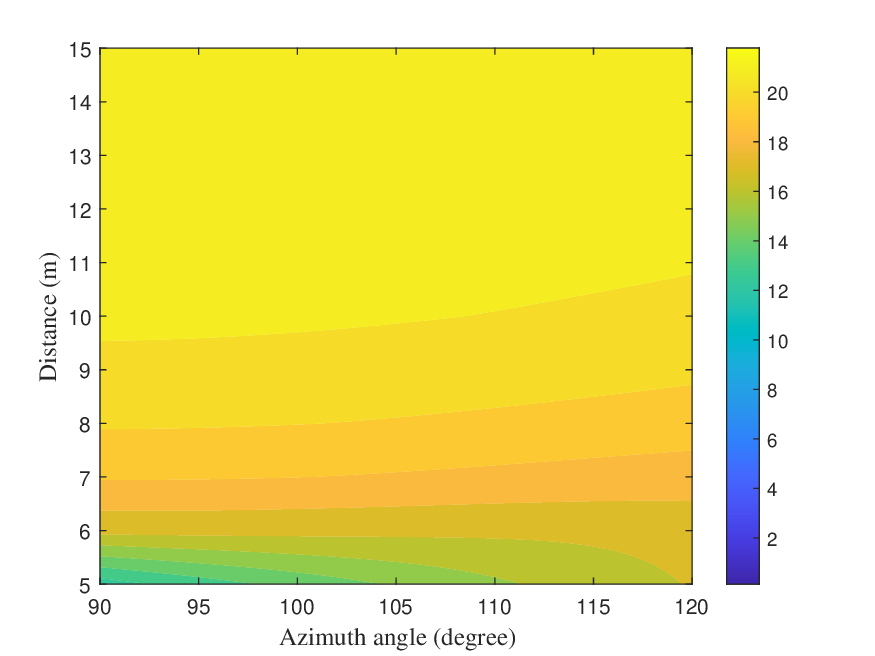}
			\end{minipage}
		}
		\subfigure[]
		{
			\begin{minipage}[b]{0.4\linewidth}
				\centering
				\includegraphics[width=1.2\textwidth, trim = 10 2 2 0,clip]{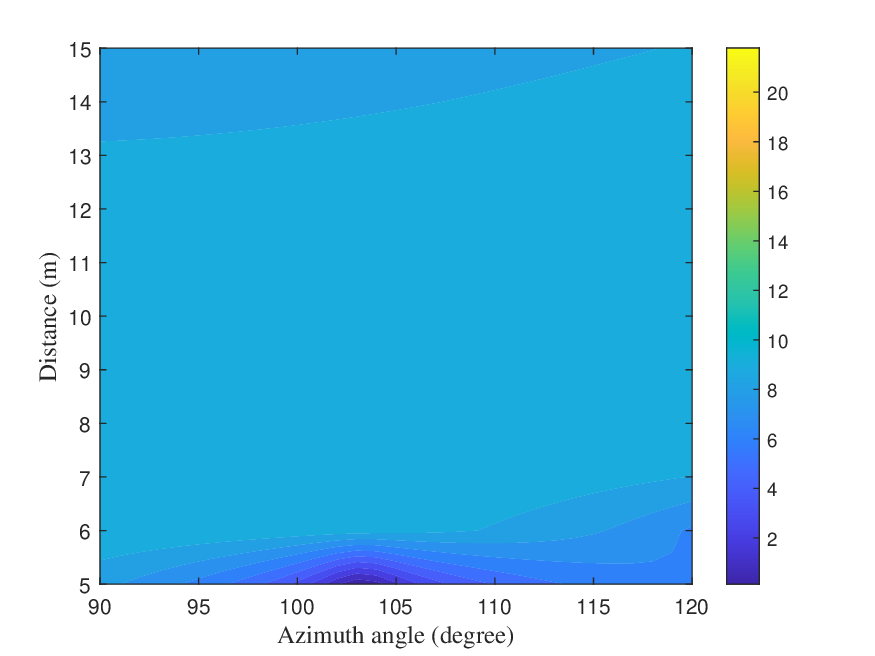}
			\end{minipage}
		}
				\subfigure[] 
		{
			\begin{minipage}[b]{0.4\linewidth}
				\centering
				\includegraphics[width=1.2\textwidth, trim = 10 2 2 0,clip]{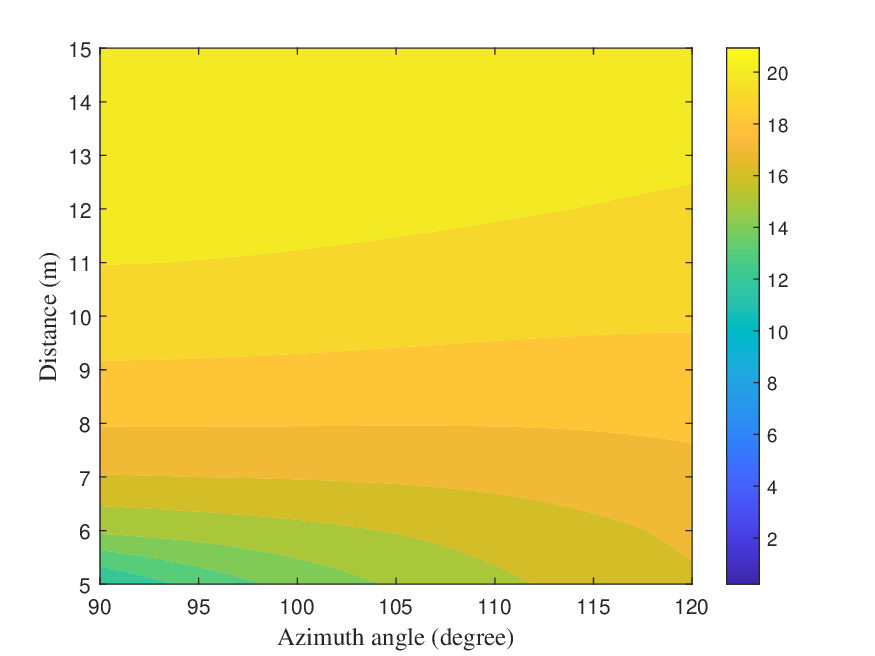}
			\end{minipage}
		}
		\subfigure[]
		{
			\begin{minipage}[b]{0.4\linewidth}
				\centering
				\includegraphics[width=1.2\textwidth, trim = 10 2 2 0,clip]{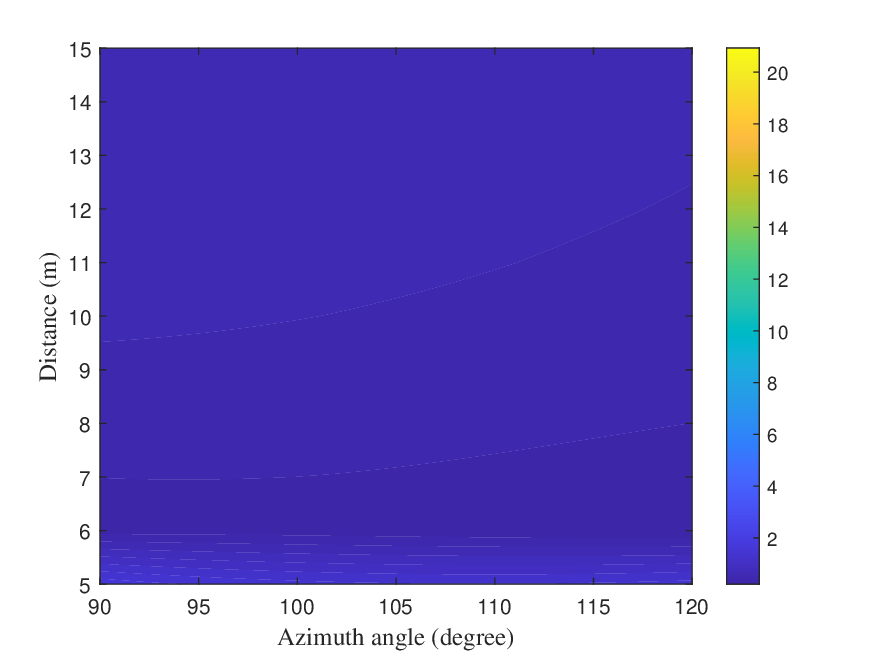}
			\end{minipage}
		}
		\caption{SR under different azimuth angles and distances from the RIS. (a) Data stream 1 under Algorithm 1. (b) Data stream 2 under Algorithm 1. (c) Data stream 1 under Algorithm 2. (d) Data stream 2 under Algorithm 2.}\label{fig4}
	\end{figure}
	
	Fig. \ref{fig4} illustrates the SR values under different azimuth angles and distances from the RIS. Due to multiplicative path loss, the SR of data stream 2 obtained using Algorithm 1 and Algorithm 2 is lower than that of data stream 1. The security performance is related to the distance between Eve and the RIS; when Eve is close to the RIS, the SR decreases. Therefore, improving the SR of data stream 2 and preventing Eve from eavesdropping on data stream 2 becomes more challenging. 
	

		\begin{figure}[t]
		\centering
		\includegraphics[width=0.45\textwidth]{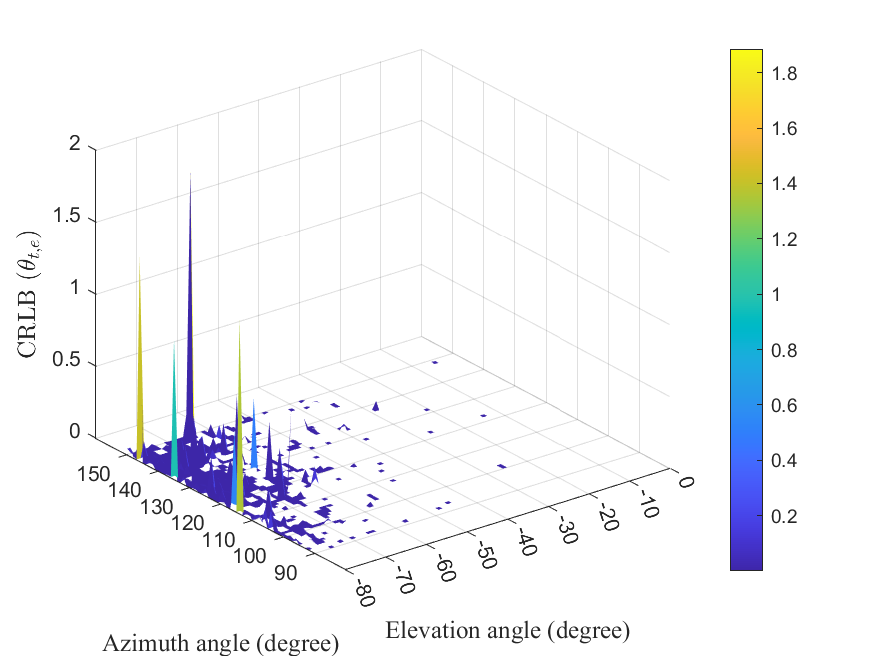}\\
		\caption{CRLB under different azimuth and elevation angles for RIS. Data stream 2 corresponding to Eve's azimuth angle.} \label{fig5_1}
	\end{figure}
	\begin{figure}[htpb]
		\centering
		\includegraphics[width=0.45\textwidth]{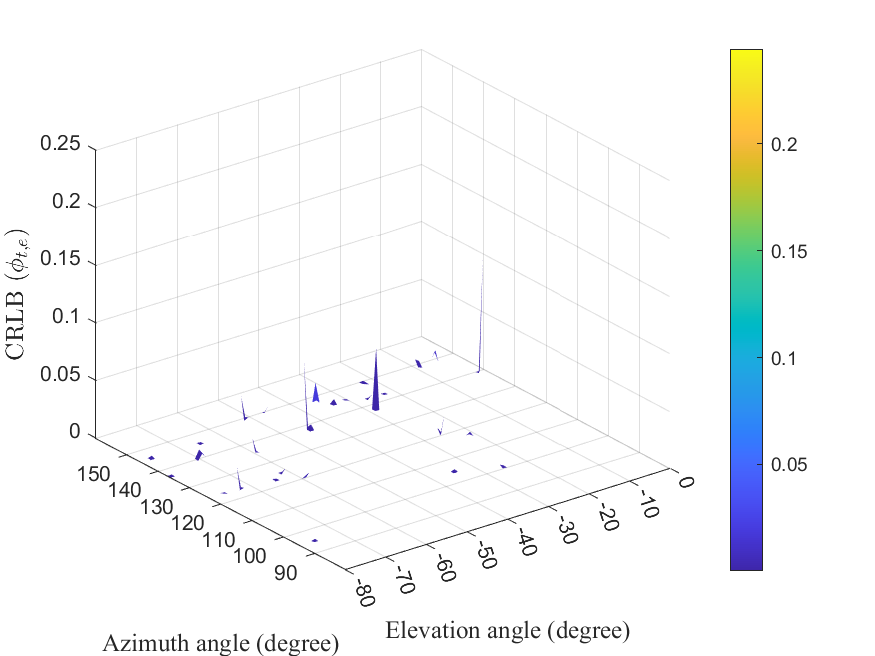}\\
		\caption{CRLB under different azimuth and elevation angles for RIS. Data stream 2 corresponding to Eve's elevation angle.  }\label{fig5_2}
	\end{figure}
\begin{figure}[t]
		\centering
		\includegraphics[width=0.45\textwidth]{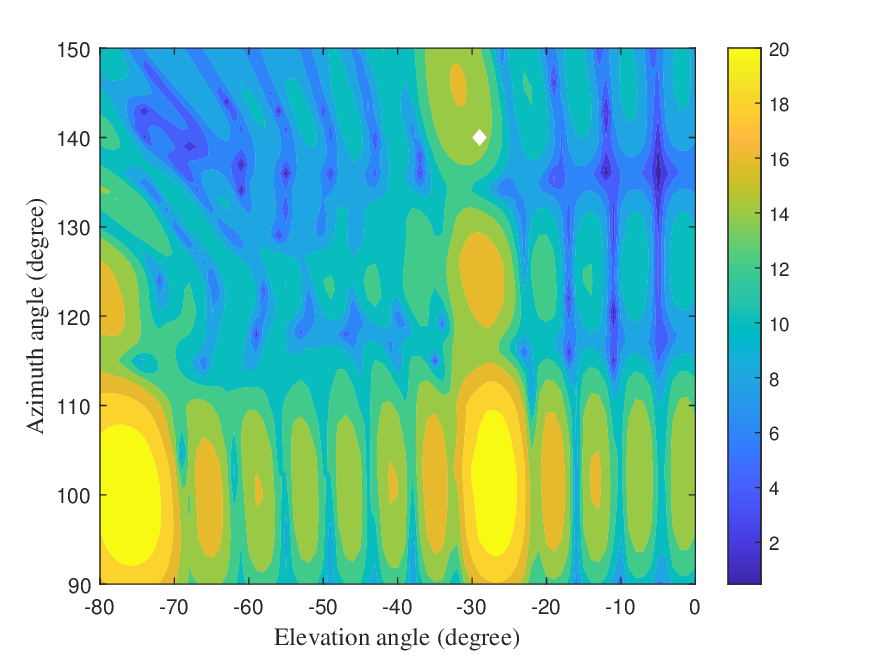}\\
		\caption{ SR versus $\tilde{\mathbf{k}}$ for Algorithm 1.}\label{6_1}
	\end{figure}
\begin{figure}[htpb]
		\centering
		\includegraphics[width=0.45\textwidth]{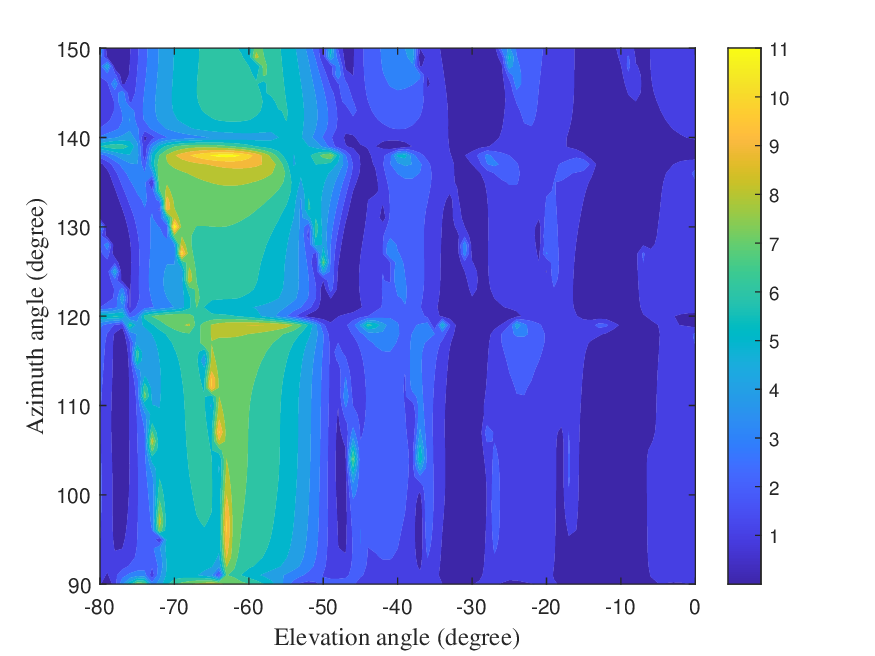}\\
		\caption{SR versus $\tilde{\mathbf{k}}$ for Algorithm 2.}\label{6_2}
	\end{figure}

	For Algorithm 1 and Algorithm 2, Figs. \ref{fig5_1} and \ref{fig5_2} show the CRLB corresponding to data stream 2. Here, the azimuth and elevation angles on the axes represent the parameters associated with $\tilde{\mathbf{k}}(\alpha,\beta)$. It can be observed that within the rotatable range for RIS, the DOA estimation performance for Eve is non-uniformly distributed, and in some directions, the estimation performance is extremely poor, which may render the designed secure beam ineffective. To address this challenge, the joint design of beamforming and DOA estimation is crucial. Moreover, when Eve is at the same height as the RIS, the rotation of the RIS becomes more sensitive to the estimation of the elevation angle, i.e., more CRLB values are high. When the elevation and azimuth angles of Eve are $0^{\circ}$ and $90^{\circ}$, respectively, Figs. \ref{6_1} and \ref{6_2} illustrate the SR distribution within the rotatable range. The optimal RIS orientations obtained using Algorithm 1 and Algorithm 2 differ. Algorithm 1 yields a more robust RIS orientation, characterized by a larger number of high SR values corresponding to $\tilde{\mathbf{k}}$, and the optional $\tilde{\mathbf{k}}$ values are continuous over a certain range. This implies that a certain delay in RIS rotation is tolerable.
	

\begin{figure}[t]
		\centering
		\includegraphics[width=0.45\textwidth]{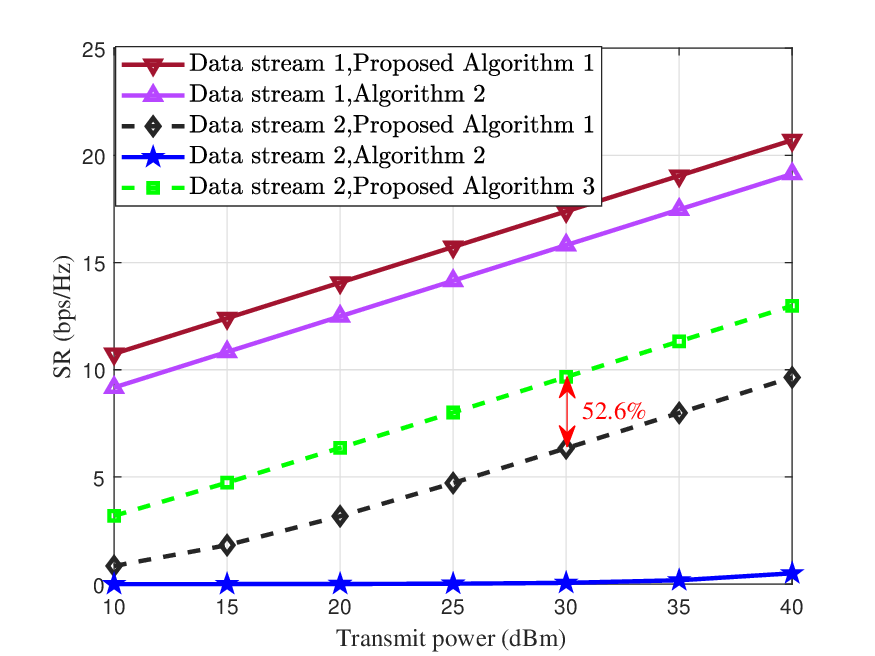}\\
		\caption{ SR versus the transmit power.}\label{TP}
	\end{figure}

	Fig. \ref{TP} illustrates the SR as a function of transmit power. As the transmit power increases, the SR improves. Although aligning the RIS toward Eve enables more accurate DOA estimation, it comes at the cost of security performance. Algorithm 3 achieves a 52.6\% improvement in SR performance compared to Algorithm 2, demonstrating that optimizing the RIS orientation through learning yields a better trade-off between DOA estimation accuracy and SR performance. Furthermore, the SR achievable by Algorithm 1 is higher than that of Algorithm 2.

\begin{figure}[t]
		\centering
		\includegraphics[width=0.45\textwidth]{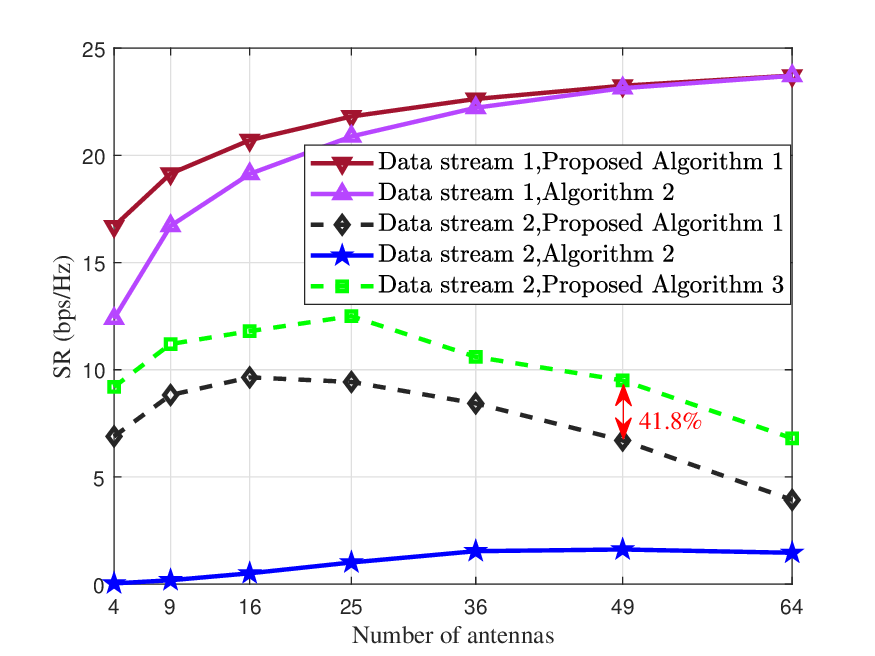}\\
		\caption{ SR versus the number of antennas.}\label{Antenna}
	\end{figure}

	The SR values under different numbers of antennas are shown in Fig. \ref{Antenna}. As the number of antennas increases, the achievable SR for data stream 1 gradually improves and then stabilizes. When the number of antennas is 49, Algorithm 1 and Algorithm 2 achieve comparable performance over the LoS path. For data stream 2, as illustrated, with an increasing number of antennas, the SR obtained using Algorithm 2 and Algorithm 3 first increases and then decreases, indicating the existence of an optimal number of antennas. Algorithm 3 achieves a 41.8\% performance improvement compared to Algorithm 2. For systems equipped with a small number of antennas, the security enhancement provided by the RIS is more significant.
	
\begin{figure}[t]
		\centering
		\includegraphics[width=0.45\textwidth]{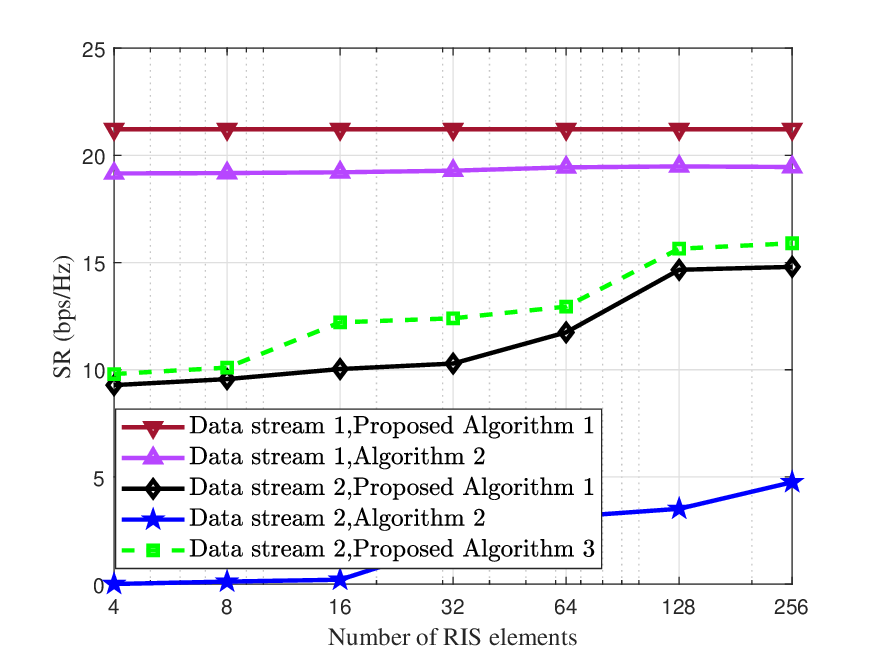}\\
		\caption{ SR versus the number of the RIS elements.}\label{RIS}
	\end{figure}

\begin{figure}[H]
		\centering
		\includegraphics[width=0.45\textwidth]{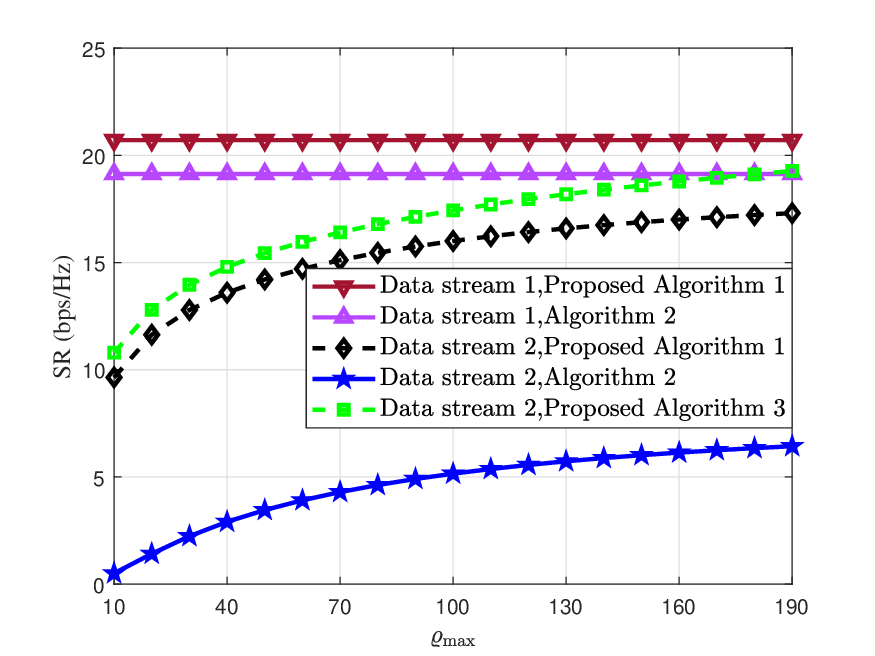}\\
		\caption{ SR versus $\varrho_{\mathrm{max}}$.}\label{rho}
	\end{figure}

	Figs. \ref{RIS} and \ref{rho} illustrate the SR variation with respect to the number of RIS elements and the maximum reflection amplitude, respectively. Increasing both the number of RIS elements and the maximum reflection amplitude contributes to further enhancing the security performance of data stream 2. A larger number of RIS elements increases the design DoF, thereby compensating for path loss. With Algorithm 3, increasing the maximum reflection amplitude can elevate the SR to a level comparable to that of data stream 1. In practice, the deployment cost of the RIS should be taken into consideration when deciding whether to deploy an RIS, as well as determining its appropriate scale and power consumption.

By jointly optimizing beamforming and RIS orientation, the proposed DSAC-T-based scheme effectively balances the trade-off between DOA estimation accuracy and security performance, a challenge that conventional optimization algorithms cannot satisfactorily resolve. The secrecy rate of the RIS-reflected data stream is substantially enhanced, especially under limited antenna configurations and severe propagation loss conditions. Moreover, optimal deployment of RIS elements and reflection coefficients is shown to provide additional degrees of freedom, enabling the system to bridge the performance gap between the LoS path and the reflected path. The robust RIS orientation obtained through reinforcement learning also accommodates certain rotation delays, thereby enhancing practicality for real-time systems. These findings indicate that a learning-driven joint design offers a promising and effective solution for realizing reliable and secure communications in future rotatable-RIS-enabled wireless networks.
	
	\section{Conclusions}\label{sec:5}
	This paper considers optimizing the orientation of an active RIS to further enhance the security performance of conventional DM networks. To address the potential degradation in DM security caused by inaccurate DoA estimation of the eavesdropper due to RIS rotation, we formulate a SR maximization problem subject to constraints on the CRLB, transmit power, rotatable range, and maximum reflection amplitude of the RIS elements. Subsequently, three algorithms are proposed. Simulation results validate that a rotatable active RIS can effectively improve SR performance. Under the framework of Algorithm 1, the feasible RIS orientations corresponding to high SR values exhibit a continuous distribution, which accommodates a certain degree of RIS rotation delay. Algorithm 3 achieves the highest SR performance among the proposed methods.
	
	\appendix
	\section{Derivation of FIM}
	With $\ln f(\hat{\mathbf{x}}_{t};\bm{\kappa})$, we can obtain
	\begin{align}
		\frac{\partial \ln f(\hat{\mathbf{x}}_{t};\bm{\kappa})}{\partial \theta_{t,e}}&=\frac{\partial \ln\frac{1}{\pi^{NP}\det(\bm{\Sigma_t})}e^{-[\hat{\mathbf{x}}_{t}-\bm{\mu}_t]^H\bm{\Sigma_t}^{-1}[\hat{\mathbf{x}}_{t}-\bm{\mu}_t]}}{\partial \theta_{t,e}}\notag\\
		&=-\frac{\partial {[\hat{\mathbf{x}}_{t}-\bm{\mu}_t]^H\bm{\Sigma_t}^{-1}[\hat{\mathbf{x}}_{t}-\bm{\mu}_t]}}{\partial \theta_{t,e}}\notag\\
		&=2\Re\left\lbrace [\hat{\mathbf{x}}_{t}-\bm{\mu}_t]^H\bm{\Sigma_t}^{-1}\mathbf{G}\frac{\partial(\mathbf{h}_{t,Re}{s}_{t,1})}{\partial \theta_{t,e}}\right\rbrace \notag\\
		&=2\Re\left\lbrace [\hat{\mathbf{x}}_{t}-\bm{\mu}_t]^H\bm{\Sigma_t}^{-1}\mathbf{G}\mathbf{b}_1\right\rbrace 
	\end{align}
	where
	\begin{align}
		&\mathbf{G}\left((p-1)N+1:pN,:\right)  = \mathbf{H}_t\mathbf{\Theta}_t^p\\
		&\mathbf{b}_1((p-1)M+1:pM) =-j\frac{2\pi{s}_{t,1}}{\lambda}\mathrm{diag}(\hat{\mathbf{b}}_1)\mathbf{h}_{t,Re}\\
		&[\hat{\mathbf{b}}_1]_m=\mathbf{i}_{t,m}^T\begin{bmatrix}
			-\sin\theta_{t,e}\cos\phi_{t,e} \\
			-\sin\theta_{t,e}\sin\phi_{t,e} \\
			\cos\theta_{t,e}
		\end{bmatrix}.
	\end{align}
	Similarly, we have
	\begin{align}
		\frac{\partial \ln f(\hat{\mathbf{x}}_{t};\bm{\kappa})}{\partial \phi_{t,e}}&=\frac{\partial \ln\frac{1}{\pi^{NP}\det(\bm{\Sigma_t})}e^{-[\hat{\mathbf{x}}_{t}-\bm{\mu}_t]^H\bm{\Sigma_t}^{-1}[\hat{\mathbf{x}}_{t}-\bm{\mu}_t]}}{\partial \phi_{t,e}}\notag\\
		&=-\frac{\partial {[\hat{\mathbf{x}}_{t}-\bm{\mu}_t]^H\bm{\Sigma_t}^{-1}[\hat{\mathbf{x}}_{t}-\bm{\mu}_t]}}{\partial \phi_{t,e}}\notag\\
		&=2\Re\left\lbrace [\hat{\mathbf{x}}_{t}-\bm{\mu}_t]^H\bm{\Sigma_t}^{-1}\mathbf{G}\frac{\partial \mathbf{h}_{t,Re}{s}_{t,1}}{\partial \phi_{t,e}}\right\rbrace \notag\\
		&=2\Re\left\lbrace [\hat{\mathbf{x}}_{t}-\bm{\mu}_t]^H\bm{\Sigma_t}^{-1}\mathbf{G}\mathbf{b}_2\right\rbrace 
	\end{align}
	where
	\begin{align}
		&\mathbf{b}_2((p-1)M+1:pM) =-j\frac{2\pi{s}_{t,1}}{\lambda}\mathrm{diag}(\hat{\mathbf{b}}_2)\mathbf{h}_{t,Re}\\
		&[\hat{\mathbf{b}}_2]_m=\mathbf{i}_{t,m}^T\begin{bmatrix}
			-\cos\theta_{t,e}\sin\phi_{t,e} \\
			\cos\theta_{t,e}\cos\phi_{t,e} \\
			0
		\end{bmatrix}.
	\end{align}
	Subsequently, the elements of the FIM matrix can be computed as
	\begin{align}
		{\Omega}_{1,1}&=\mathbb{E}\{\frac{\partial \ln^H f(\hat{\mathbf{x}}_{t};\bm{\kappa})}{\partial \theta_{t,e}}\frac{\partial \ln f(\hat{\mathbf{x}}_{t};\bm{\kappa})}{\partial \theta_{t,e}}\}\notag\\
		&=\mathbb{E}\{\hat{\sigma}_t^{-2}(\mathbf{b}_1^H\mathbf{G}^H[\hat{\mathbf{x}}_{t}-\bm{\mu}_t]+\mathbf{b}_1^T\mathbf{G}^T[\hat{\mathbf{x}}_{t}-\bm{\mu}_t]^*)\notag\\
		&\quad\ \hat{\sigma}_t^{-2}([\hat{\mathbf{x}}_{t}-\bm{\mu}_t]^H\mathbf{G}\mathbf{b}_1+[\hat{\mathbf{x}}_{t}-\bm{\mu}_t]^T\mathbf{G}^*\mathbf{b}_1^*)\}\notag\\
		&=2\hat{\sigma}_t^{-2}{\mathbf{b}_1^H\mathbf{G}^H\mathbf{G}\mathbf{b}_1}\\
		{\Omega}_{2,2}&=\mathbb{E}\{\frac{\partial \ln^H f(\hat{\mathbf{x}}_{t};\bm{\kappa})}{\partial \phi_{t,e}}\frac{\partial \ln f(\hat{\mathbf{x}}_{t};\bm{\kappa})}{\partial \phi_{t,e}}\}\notag\\
		&=\mathbb{E}\{\hat{\sigma}_t^{-2}(\mathbf{b}_2^H\mathbf{G}^H[\hat{\mathbf{x}}_{t}-\bm{\mu}_t]+\mathbf{b}_2^T\mathbf{G}^T[\hat{\mathbf{x}}_{t}-\bm{\mu}_t]^*)\notag\\
		&\quad\ \hat{\sigma}_t^{-2}([\hat{\mathbf{x}}_{t}-\bm{\mu}_t]^H\mathbf{G}\mathbf{b}_2+[\hat{\mathbf{x}}_{t}-\bm{\mu}_t]^T\mathbf{G}^*\mathbf{b}_2^*)\}\notag\\
		&=2\hat{\sigma}_t^{-2}{\mathbf{b}_2^H\mathbf{G}^H\mathbf{G}\mathbf{b}_2}\\
		{\Omega}_{1,2}&=\mathbb{E}\{\frac{\partial \ln^H f(\hat{\mathbf{x}}_{t};\bm{\kappa})}{\partial \theta_{t,e}}\frac{\partial \ln f(\hat{\mathbf{x}}_{t};\bm{\kappa})}{\partial \phi_{t,e}}\}\notag\\
		&=\mathbb{E}\{\hat{\sigma}_t^{-2}(\mathbf{b}_1^H\mathbf{G}^H[\hat{\mathbf{x}}_{t}-\bm{\mu}_t]+\mathbf{b}_1^T\mathbf{G}^T[\hat{\mathbf{x}}_{t}-\bm{\mu}_t]^*)\notag\\
		&\quad\ \hat{\sigma}_t^{-2}([\hat{\mathbf{x}}_{t}-\bm{\mu}_t]^H\mathbf{G}\mathbf{b}_2+[\hat{\mathbf{x}}_{t}-\bm{\mu}_t]^T\mathbf{G}^*\mathbf{b}_2^*)\}\notag\\
		&=2\hat{\sigma}_t^{-2}\Re\{\mathbf{b}_1^H\mathbf{G}^H\mathbf{G}\mathbf{b}_2\}\\
		{\Omega}_{2,1}&=\mathbb{E}\{\frac{\partial \ln^H f(\hat{\mathbf{x}}_{t};\bm{\kappa})}{\partial \phi_{t,e}}\frac{\partial \ln f(\hat{\mathbf{x}}_{t};\bm{\kappa})}{\partial \theta_{t,e}}\}\notag\\
		&=\mathbb{E}\{\hat{\sigma}_t^{-2}(\mathbf{b}_2^H\mathbf{G}^H[\hat{\mathbf{x}}_{t}-\bm{\mu}_t]+\mathbf{b}_2^T\mathbf{G}^T[\hat{\mathbf{x}}_{t}-\bm{\mu}_t]^*)\notag\\
		&\quad\ \hat{\sigma}_t^{-2}([\hat{\mathbf{x}}_{t}-\bm{\mu}_t]^H\mathbf{G}\mathbf{b}_1+[\hat{\mathbf{x}}_{t}-\bm{\mu}_t]^T\mathbf{G}^*\mathbf{b}_1^*)\}\notag\\
		&=2\hat{\sigma}_t^{-2}\Re\{\mathbf{b}_2^H\mathbf{G}^H\mathbf{G}\mathbf{b}_1\}.
	\end{align}

\section*{Acknowledgements}
This work was supported in part by Hainan Provincial Natural Science Foundation of China under Grant 626ZD0993, Grant 526QN0542, and Grant 626QN0553; in part by the National Natural Science Foundation of China under Grant U22A2002; in part by the Hainan Province Science and Technology Special Fund under Grant ZDYF2024GXJS292; and in part by the National Key Research and Development Program of China under Grant 2023YFF0612900.

%


\bibliographystyle{elsarticle-num}
\balance
\bibliography{mybib}

\begin{thebibliography}{10}
\expandafter\ifx\csname url\endcsname\relax
\def\url#1{\texttt{#1}}\fi
\expandafter\ifx\csname urlprefix\endcsname\relax\def\urlprefix{URL }\fi
\expandafter\ifx\csname href\endcsname\relax
\def\href#1#2{#2} \def\path#1{#1}\fi

\bibitem{Xie2025}
W.~Xie, Z.~Li, C.~Yu, H.~Xu, J.~Wang, W.~Wu, X.~Li, L.~Yang,
Movable-antenna-assisted covert communications with reconfigurable
intelligent surfaces, \textit{IEEE Internet Things J.} 12~(9) (2025)
12369--12382.
\newblock \href {http://dx.doi.org/10.1109/jiot.2024.3520710}
{\path{doi:10.1109/jiot.2024.3520710}}.

\bibitem{Wong2022}
K.-K. Wong, K.-F. Tong, Fluid antenna multiple access, \textit{IEEE Trans. Wireless Commun.} 21~(7) (2022) 4801--4815.
\newblock \href {http://dx.doi.org/10.1109/TWC.2021.3133410}
{\path{doi:10.1109/TWC.2021.3133410}}.

\bibitem{Chen2025b}
J.~Chen, Y.~Xiao, Z.~Peng, J.~Zhu, X.~Lei, C.~Masouros, K.-K. Wong, Hybrid
beamforming for RIS-assisted multiuser fluid antenna systems, \textit{IEEE Trans. Wireless Commun.} (2025) 1--1\href
{http://dx.doi.org/10.1109/TWC.2025.3598493}
{\path{doi:10.1109/TWC.2025.3598493}}.

\bibitem{Wong2023}
K.-K. Wong, D.~Morales-Jimenez, K.-F. Tong, C.-B. Chae, Slow fluid antenna
multiple access, \textit{IEEE Trans. Commun.} 71~(5) (2023)
2831--2846.
\newblock \href {http://dx.doi.org/10.1109/TCOMM.2023.3255904}
{\path{doi:10.1109/TCOMM.2023.3255904}}.

\bibitem{Bian2026}
H.~Bian, F.~Zhao, M.~Chen, L.~Liu, Y.~Yao, H.~Jiang, F.~Shu, Movable
antenna-enabled secure transmission for active RIS-aided ISAC systems, \textit{IEEE Trans. Netw. Sci. Eng.} 13 (2026) 6325--6344.
\newblock \href {http://dx.doi.org/10.1109/TNSE.2026.3657852}
{\path{doi:10.1109/TNSE.2026.3657852}}.

\bibitem{Tang2025}
J.~Tang, C.~Pan, Y.~Zhang, H.~Ren, K.~Wang, Secure MIMO communication relying
on movable antennas, \textit{IEEE Trans. Commun.} 73~(4) (2025)
2159--2175.
\newblock \href {http://dx.doi.org/10.1109/tcomm.2024.3465369}
{\path{doi:10.1109/tcomm.2024.3465369}}.

\bibitem{Li2025c}
M.~Li, F.~Shu, Y.~Si, R.~Chen, C.~Pan, Y.~Wu, Near-field directional modulation
for RIS-aided movable antenna MIMO systems with hardware impairments, \textit{IEEE Trans. Netw. Sci. Eng.} (2025) 1--16\href
{http://dx.doi.org/10.1109/TNSE.2025.3594342}
{\path{doi:10.1109/TNSE.2025.3594342}}.

\bibitem{Ding2025a}
Z.~Ding, R.~Schober, H.~Vincent~Poor, Flexible-antenna systems: A
pinching-antenna perspective, \textit{IEEE Trans. Commun.} 73~(10)
(2025) 9236--9253.
\newblock \href {http://dx.doi.org/10.1109/TCOMM.2025.3555866}
{\path{doi:10.1109/TCOMM.2025.3555866}}.

\bibitem{Wang2025a}
K.~Wang, Z.~Ding, G.~K. Karagiannidis, Antenna activation and resource
allocation in multi-waveguide pinching-antenna systems, \textit{IEEE Trans. Wireless Commun.} (2025) 1--1\href
{http://dx.doi.org/10.1109/TWC.2025.3608068}
{\path{doi:10.1109/TWC.2025.3608068}}.

\bibitem{Xu2025}
X.~Xu, X.~Mu, Z.~Wang, Y.~Liu, A.~Nallanathan, Pinching-antenna systems (pass):
Power radiation model and optimal beamforming design, \textit{IEEE Trans. Commun.} (2025) 1--1\href
{http://dx.doi.org/10.1109/TCOMM.2025.3636083}
{\path{doi:10.1109/TCOMM.2025.3636083}}.

\bibitem{Qu2025}
K.~Qu, H.~Li, C.~Sun, W.~Zhang, S.~Guo, H.~Zhang, Rotatable array-enabled
multi-bs cooperative ISAC transmit beampattern design, \textit{IEEE Trans. Veh. Technol.} 74~(9) (2025) 14775--14780.
\newblock \href {http://dx.doi.org/10.1109/TVT.2025.3559665}
{\path{doi:10.1109/TVT.2025.3559665}}.

\bibitem{Zeng2020}
Q.~Zeng, Z.~Xue, W.~Ren, W.~Li, Dual-band beam-scanning antenna using rotatable
planar phase gradient transmitarrays, \textit{IEEE Trans. Antennas Propag.} 68~(6) (2020) 5021--5026.
\newblock \href {http://dx.doi.org/10.1109/TAP.2020.2963929}
{\path{doi:10.1109/TAP.2020.2963929}}.

\bibitem{Zheng2025}
B.~Zheng, T.~Ma, C.~You, J.~Tang, R.~Schober, R.~Zhang, Rotatable antenna
enabled wireless communication and sensing: Opportunities and challenges,
\textit{IEEE Wireless Commun.} (2025) 1--8\href
{http://dx.doi.org/10.1109/MWC.2025.3611919}
{\path{doi:10.1109/MWC.2025.3611919}}.

\bibitem{Shu2024}
F.~Shu, Y.~Wang, X.~Wang, G.~Xia, L.~Yang, W.~Shi, C.~Shen, J.~Wang, Precoding
and beamforming design for intelligent reconfigurable surface-aided hybrid
secure spatial modulation, \textit{IEEE Trans. Wireless Commun.}
23~(9) (2024) 11770--11784.
\newblock \href {http://dx.doi.org/10.1109/TWC.2024.3384969}
{\path{doi:10.1109/TWC.2024.3384969}}.

\bibitem{Man2025}
Y.~Man, P.~Yang, L.~Yin, H.~Yang, Y.~Zhao, Z.~Wu, Y.~Pu, Z.~Wen, Y.~Luo, An
efficient multibeamforming method based on 1-bit phase modulation for
time-modulated arrays, \textit{IEEE Trans. Antennas Propag.} 73~(6)
(2025) 3654--3665.
\newblock \href {http://dx.doi.org/10.1109/tap.2025.3551614}
{\path{doi:10.1109/tap.2025.3551614}}.

\bibitem{Shu2018}
F.~Shu, X.~Wu, J.~Hu, J.~Li, R.~Chen, J.~Wang, Secure and precise wireless
transmission for random-subcarrier-selection-based directional modulation
transmit antenna array, \textit{IEEE J. Sel. Areas Commun.}
36~(4) (2018) 890--904.
\newblock \href {http://dx.doi.org/10.1109/JSAC.2018.2824231}
{\path{doi:10.1109/JSAC.2018.2824231}}.

\bibitem{Fang2024}
X.~Fang, M.~Li, S.~Li, D.~Ramaccia, A.~Toscano, F.~Bilotti, D.~Ding, Diverse
frequency time modulation for passive false target spoofing: Design and
experiment, \textit{IEEE Trans. Microw. Theory Techn.} 72~(3)
(2024) 1932--1942.
\newblock \href {http://dx.doi.org/10.1109/TMTT.2023.3305187}
{\path{doi:10.1109/TMTT.2023.3305187}}.

\bibitem{Li2025}
H.~Li, Y.~Chen, S.~Yang, Independent control of multiple harmonic beams in
time-modulated arrays with subarrayed time-segmented pseudorandom modulation,
\textit{IEEE Trans. Antennas Propag.} 73~(8) (2025) 5559--5573.
\newblock \href {http://dx.doi.org/10.1109/tap.2025.3562928}
{\path{doi:10.1109/tap.2025.3562928}}.

\bibitem{Chen2025}
H.~Chen, J.~Li, S.~Yang, W.~Liu, Y.~C. Eldar, C.~Yuen, Near-field source
localization in 3-D using two parallel centrally symmetric unfold coprime
array, \textit{IEEE Trans. Wireless Commun.} 24~(6) (2025) 4738--4749.
\newblock \href {http://dx.doi.org/10.1109/TWC.2025.3543616}
{\path{doi:10.1109/TWC.2025.3543616}}.

\bibitem{Zhuang2024}
Y.~Zhuang, X.~Zhang, Z.~He, M.~S. Greco, F.~Gini, Sparse array design via
integer linear programming, \textit{IEEE Trans. Signal Process.} 72 (2024)
4812--4826.
\newblock \href {http://dx.doi.org/10.1109/TSP.2024.3460383}
{\path{doi:10.1109/TSP.2024.3460383}}.

\bibitem{Shi2024}
W.~Shi, Q.~Wu, D.~Wu, F.~Shu, J.~Wang, Joint transmit and reflective
beamforming design for active IRS-aided swipt systems, \textit{Chin. J. Electron.} 33~(2) (2024) 536--548.
\newblock \href {http://dx.doi.org/10.23919/cje.2022.00.287}
{\path{doi:10.23919/cje.2022.00.287}}.

\bibitem{Chen2025a}
W.~Chen, J.~Luo, H.~Ding, S.~Wang, F.~Gong, Three-dimensional fluid
antenna-assisted covert communications with friendly jamming, \textit{IEEE Trans. Wireless Commun.} (2025) 1--1\href
{http://dx.doi.org/10.1109/TWC.2025.3609276}
{\path{doi:10.1109/TWC.2025.3609276}}.

\bibitem{Zhang2025a}
Q.~Zhang, W.~Gao, C.~Liu, Y.~Yao, S.~Yan, F.~Shu, S.~Jin, Covert transmission
for active RIS-aided full-duplex uav integrated sensing, communication, and
computation systems, \textit{IEEE J. Sel. Areas Commun.} (2025)
1--1\href {http://dx.doi.org/10.1109/JSAC.2025.3637785}
{\path{doi:10.1109/JSAC.2025.3637785}}.

\bibitem{Li2025a}
M.~Li, J.~Xin, F.~Shu, X.~Wang, Y.~Wu, C.~Pan, Secure directional modulation
with movable antenna array aided by RIS, \textit{IEEE Trans. Veh. Technol.} (2025) 1--14\href {http://dx.doi.org/10.1109/tvt.2025.3634592}
{\path{doi:10.1109/tvt.2025.3634592}}.

\bibitem{Li2026}
M.~Li, W.~Gao, Q.~Wu, F.~Shu, C.~Pan, D.~Wu, Direction modulation design for
uav assisted by IRS with discrete phase shift, \textit{IEEE Trans. Green Commun. Netw.} 10 (2026) 172--186.
\newblock \href {http://dx.doi.org/10.1109/TGCN.2025.3572113}
{\path{doi:10.1109/TGCN.2025.3572113}}.

\bibitem{Ma2025}
Y.~Ma, K.~Liu, Y.~Liu, L.~Zhu, Movable antenna empowered secure near-field MIMO
communications, \textit{IEEE Trans. Commun.} (2025) 1--1\href
{http://dx.doi.org/10.1109/tcomm.2025.3612581}
{\path{doi:10.1109/tcomm.2025.3612581}}.

\bibitem{Li2025b}
K.~Li, K.~Yu, D.~Ma, Y.~Zhao, X.~Liu, Q.~Zhang, Z.~Feng, Can movable
antenna-enabled micro-mobility replace uav-enabled macro-mobility? a physical
layer security perspective, \textit{IEEE Trans. Mobile Comput.} (2025)
1--13\href {http://dx.doi.org/10.1109/tmc.2025.3624340}
{\path{doi:10.1109/tmc.2025.3624340}}.

\bibitem{Hu2024}
G.~Hu, Q.~Wu, D.~Xu, K.~Xu, J.~Si, Y.~Cai, N.~Al-Dhahir, Movable
antennas-assisted secure transmission without eavesdroppers’ instantaneous
csi, \textit{IEEE Trans. Mobile Comput.} 23~(12) (2024) 14263--14279.
\newblock \href {http://dx.doi.org/10.1109/tmc.2024.3438795}
{\path{doi:10.1109/tmc.2024.3438795}}.

\bibitem{Ding2025}
J.~Ding, Z.~Zhou, B.~Jiao, Movable antenna-aided secure full-duplex multi-user
communications, \textit{IEEE Trans. Wireless Commun.} 24~(3) (2025)
2389--2403.
\newblock \href {http://dx.doi.org/10.1109/twc.2024.3520806}
{\path{doi:10.1109/twc.2024.3520806}}.

\bibitem{Ma2025a}
Y.~Ma, K.~Liu, Y.~Liu, L.~Zhu, Z.~Xiao, Movable-antenna aided secure
transmission for RIS-ISAC systems, \textit{IEEE Trans. Wireless Commun.} (2025) 1--1\href {http://dx.doi.org/10.1109/twc.2025.3577040}
{\path{doi:10.1109/twc.2025.3577040}}.

\bibitem{LeHung2025}
H.~Le~Hung, N.~H. Huy, N.~C. Luong, Q.-V. Pham, D.~Niyato, N.~T. Hoa,
Beamforming design for physical security in movable antenna-aided ISAC
systems: A reinforcement learning approach, \textit{IEEE Trans. Veh. Technol.} 74~(11) (2025) 18163--18167.
\newblock \href {http://dx.doi.org/10.1109/tvt.2025.3575653}
{\path{doi:10.1109/tvt.2025.3575653}}.

\bibitem{Wei2025}
X.~Wei, W.~Mei, Q.~Wu, Q.~Jia, B.~Ning, Z.~Chen, J.~Fang, Movable antennas meet
intelligent reflecting surface: Friends or foes?, \textit{IEEE Trans. Commun.} 73~(11) (2025) 12756--12770.
\newblock \href {http://dx.doi.org/10.1109/tcomm.2025.3588579}
{\path{doi:10.1109/tcomm.2025.3588579}}.

\bibitem{Zhang2025}
X.~Zhang, L.~Xiang, J.~Wang, X.~Gao, D.~W.~K. Ng, R.~Schober, Rotatable antenna
array enabled uav mmwave massive MIMO communication, \textit{IEEE Trans. Commun.} (2025) 1--1\href
{http://dx.doi.org/10.1109/tcomm.2025.3622962}
{\path{doi:10.1109/tcomm.2025.3622962}}.

\bibitem{Xiong2025}
Y.~Xiong, S.~Yang, S.~Sun, L.~Liu, Z.~Zhang, N.~Wei, Rotatable and movable
antenna enhanced multiuser communications: Rotation and position
optimization, \textit{IEEE Internet Things J.} 12~(24) (2025) 55822--55837.
\newblock \href {http://dx.doi.org/10.1109/jiot.2025.3626517}
{\path{doi:10.1109/jiot.2025.3626517}}.

\bibitem{Xiong2025a}
X.~Xiong, B.~Zheng, W.~Wu, W.~Zhu, M.~Wen, S.~Lin, Y.~Zeng, Intelligent
rotatable antenna for integrated sensing, communication, and computation:
Challenges and opportunities, \textit{IEEE Wireless Commun.} (2025) 1--8\href
{http://dx.doi.org/10.1109/mwc.2025.3622912}
{\path{doi:10.1109/mwc.2025.3622912}}.

\bibitem{Zheng2025a}
B.~Zheng, Q.~Wu, T.~Ma, R.~Zhang,
\href{https://arxiv.org/abs/2501.02595}{Rotatable antenna enabled wireless
communication: Modeling and optimization} (2025).
\newblock \href {http://arxiv.org/abs/2501.02595} {\path{arXiv:2501.02595}}.
\newline\urlprefix\url{https://arxiv.org/abs/2501.02595}

\bibitem{Zheng2025b}
B.~Zheng, T.~Ma, C.~You, J.~Tang, R.~Schober, R.~Zhang,
\href{https://arxiv.org/abs/2505.16828}{Rotatable antenna enabled wireless
communication and sensing: Opportunities and challenges} (2025).
\newblock \href {http://arxiv.org/abs/2505.16828} {\path{arXiv:2505.16828}}.
\newline\urlprefix\url{https://arxiv.org/abs/2505.16828}

\bibitem{Dai2025}
L.~Dai, B.~Zheng, Q.~Wu, C.~You, R.~Schober, R.~Zhang, Rotatable
antenna-enabled secure wireless communication, \textit{IEEE Wireless Commun. Lett.} 14~(11) (2025) 3440--3444.
\newblock \href {http://dx.doi.org/10.1109/lwc.2025.3593258}
{\path{doi:10.1109/lwc.2025.3593258}}.

\bibitem{Shao2025a}
X.~Shao, Q.~Jiang, R.~Zhang, 6D movable antenna based on user distribution:
Modeling and optimization, \textit{IEEE Trans. Wireless Commun.}
24~(1) (2025) 355--370.
\newblock \href {http://dx.doi.org/10.1109/twc.2024.3492195}
{\path{doi:10.1109/twc.2024.3492195}}.

\bibitem{Shao2025}
X.~Shao, R.~Zhang, Q.~Jiang, R.~Schober, 6d movable antenna enhanced wireless
network via discrete position and rotation optimization, \textit{IEEE J. Sel. Areas Commun.} 43~(3) (2025) 674--687.
\newblock \href {http://dx.doi.org/10.1109/JSAC.2025.3531571}
{\path{doi:10.1109/JSAC.2025.3531571}}.

\bibitem{Shao2025b}
X.~Shao, W.~Mei, C.~You, Q.~Wu, B.~Zheng, C.-X. Wang, J.~Li, R.~Zhang,
R.~Schober, L.~Zhu, W.~Zhuang, X.~Shen, A tutorial on six-dimensional movable
antenna for 6G networks: Synergizing positionable and rotatable antennas,
\textit{IEEE Commun. Surv. Tutorials} (2025) 1--1\href
{http://dx.doi.org/10.1109/comst.2025.3602939}
{\path{doi:10.1109/comst.2025.3602939}}.

\bibitem{Shao2025d}
X.~Shao, L.~Hu, Y.~Sun, X.~Li, Y.~Zhang, J.~Ding, X.~Shi, F.~Chen, D.~W.~K. Ng,
R.~Schober, Hybrid near-far field 6d movable antenna design exploiting
directional sparsity and deep learning, \textit{IEEE Trans. Wireless Commun.} (2025) 1--1\href {http://dx.doi.org/10.1109/TWC.2025.3605550}
{\path{doi:10.1109/TWC.2025.3605550}}.

\bibitem{Wang2025}
W.~Wang, Y.~Huang, X.~Shao, C.~Zhang, Aerial 6D movable antenna-enabled
cell-free networks, \textit{IEEE Trans. Veh. Technol.} (2025)
1--6\href {http://dx.doi.org/10.1109/tvt.2025.3614719}
{\path{doi:10.1109/tvt.2025.3614719}}.

\bibitem{Pi2025}
X.~Pi, L.~Zhu, H.~Mao, Z.~Xiao, X.-G. Xia, R.~Zhang, 6d movable antenna
enhanced multi-access point coordination via position and orientation
optimization, \textit{IEEE Trans. Wireless Commun.} (2025) 1--1\href
{http://dx.doi.org/10.1109/twc.2025.3587803}
{\path{doi:10.1109/twc.2025.3587803}}.

\bibitem{Shao2025c}
X.~Shao, R.~Zhang, Q.~Jiang, J.~Park, T.~Q.~S. Quek, R.~Schober, Distributed
channel estimation and optimization for 6d movable antenna: Unveiling
directional sparsity, \textit{IEEE J. Sel. Topics Signal Process.}
19~(2) (2025) 349--365.
\newblock \href {http://dx.doi.org/10.1109/jstsp.2025.3539085}
{\path{doi:10.1109/jstsp.2025.3539085}}.

\bibitem{Hua2025}
H.~Hua, Y.~Zhou, W.~Mei, J.~Xu, R.~Zhang, Hierarchically tunable 6DMA for
wireless communication and sensing: Modeling and performance optimization,
\textit{IEEE Trans. Wireless Commun.} (2025) 1--1\href
{http://dx.doi.org/10.1109/twc.2025.3613548}
{\path{doi:10.1109/twc.2025.3613548}}.

\bibitem{Lu2025}
H.~Lu, Z.~Yu, Y.~Zeng, S.~Ma, S.~Jin, R.~Zhang, Wireless communication with
flexible reflector: Joint placement and rotation optimization for coverage
enhancement, \textit{IEEE Trans. Wireless Commun.} 24~(10) (2025)
8252--8266.
\newblock \href {http://dx.doi.org/10.1109/twc.2025.3564956}
{\path{doi:10.1109/twc.2025.3564956}}.

\bibitem{Wang2025c}
K.~Wang, C.~Ouyang, Y.~Liu, Z.~Ding, Pinching-antenna systems with LoS
blockages, \textit{IEEE Wireless Commun. Lett.} 14~(12) (2025) 4122--4126.
\newblock \href {http://dx.doi.org/10.1109/lwc.2025.3614451}
{\path{doi:10.1109/lwc.2025.3614451}}.

\bibitem{Wang2025b}
K.~Wang, Z.~Ding, N.~Al-Dhahir, Pinching-antenna systems for physical layer
security, \textit{IEEE Wireless Commun. Lett.} (2025) 1--1\href
{http://dx.doi.org/10.1109/LWC.2025.3624885}
{\path{doi:10.1109/LWC.2025.3624885}}.

\bibitem{Chen2024}
Z.~Chen, P.~Chen, Z.~Guo, Y.~Zhang, X.~Wang, A RIS-based vehicle DoA estimation
method with integrated sensing and communication system, \textit{IEEE Trans. Intell. Transp. Syst.} 25~(6) (2024) 5554--5566.
\newblock \href {http://dx.doi.org/10.1109/TITS.2023.3330172}
{\path{doi:10.1109/TITS.2023.3330172}}.

\bibitem{Duan2022}
J.~Duan, Y.~Guan, S.~E. Li, Y.~Ren, Q.~Sun, B.~Cheng, Distributional soft
actor-critic: Off-policy reinforcement learning for addressing value
estimation errors, \textit{IEEE Trans. Neural Netw. Learn. Syst.}
33~(11) (2022) 6584--6598.
\newblock \href {http://dx.doi.org/10.1109/TNNLS.2021.3082568}
{\path{doi:10.1109/TNNLS.2021.3082568}}.

\bibitem{Duan2025}
J.~Duan, W.~Wang, L.~Xiao, J.~Gao, S.~E. Li, C.~Liu, Y.-Q. Zhang, B.~Cheng,
K.~Li, Distributional soft actor-critic with three refinements, \textit{IEEE Trans. Pattern Anal. Mach. Intell.} 47~(5) (2025)
3935--3946.
\newblock \href {http://dx.doi.org/10.1109/TPAMI.2025.3537087}
{\path{doi:10.1109/TPAMI.2025.3537087}}.

\end{thebibliography}
~~~\\
~~~\\







%

\end{document}